\documentclass[showpacs,aps,prd,reprint,superscriptaddress,nofootinbib,longbibliography,preprintnumbers]{revtex4-1}
\usepackage[colorlinks=true, pdfstartview=FitV, linkcolor=magenta,citecolor=blue, urlcolor=magenta,
bookmarks=true, bookmarksnumbered=true, breaklinks]{hyperref}
\usepackage[dvipdfmx]{graphicx}
\usepackage{here}
\usepackage{url}
\usepackage{amsmath,amssymb,bm,color,longtable,mathrsfs,amsfonts,slashed}

\begin{document}
\begin{flushright}
\end{flushright}

\preprint{YITP-22-119, J-PARC-TH-0279}

\title{Effective model for pure Yang-Mills theory on $\mathbb{T}^2\times \mathbb{R}^2$ with Polyakov loops}

\author{Daiki~Suenaga}
\email[]{daiki.suenaga@riken.jp}
\affiliation{Strangeness Nuclear Physics Laboratory, RIKEN Nishina Center, Wako 351-0198, Japan}
\affiliation{Research Center for Nuclear Physics,
Osaka University, Ibaraki 567-0048, Japan }

\author{Masakiyo~Kitazawa}
\email[]{kitazawa@yukawa.kyoto-u.ac.jp}
\affiliation{Yukawa Institute for Theoretical Physics, Kyoto University, Kyoto, 606-8502 Japan}
\affiliation{J-PARC Branch, KEK Theory Center, Institute of Particle and Nuclear Studies, KEK, 319-1106 Japan}
\affiliation{Department of Physics, Osaka University, Toyonaka, Osaka, 560-0043 Japan}

\date{\today}

\begin{abstract}
We investigate the phase diagram and thermodynamics of $SU(N)$ pure Yang-Mills theory on a manifold $\mathbb{T}^2\times \mathbb{R}^2$ with an effective model that includes two Polyakov loops along two compactified directions. We find that a rich phase structure can appear owing to the spontaneous breaking of two center symmetries for $N=2$ and $3$. Thermodynamic quantities are obtained in the model and compared with recent lattice results. It is shown that two Polyakov loops play significant roles in thermodynamics on $\mathbb{T}^2\times \mathbb{R}^2$.
\end{abstract}


\maketitle

\section{Introduction}
\label{sec:Introduction}

Boundary conditions (BCs) in quantum field theory give rise to nontrivial modifications of the system. A renowned example is the Casimir effect in quantum electrodynamics~\cite{Casimir:1948dh}, where the effect of the BC imposed by conductors is observable as the force acting on them~\cite{Lamoreaux:1996wh,Mohideen:1998iz}. In the Matsubara formalism for thermal field theory, nonzero temperature ($T$) is introduced as the BC along the imaginary-time direction in the Euclidean spacetime. Various phenomena in thermal systems, such as phase transitions, thus can be viewed as those provoked by the BC.

In $SU(N)$ Yang-Mills (YM) theory, a BC introduces a new symmetry of the action, that is the center symmetry of the gauge group, $Z_N$, through the twist at the BC. YM theory at nonzero $T$ thus has a $Z_N$ symmetry. The deconfinement phase transition at nonzero $T$ is characterized by the spontaneous breaking of this symmetry in the high temperature phase~\cite{Svetitsky:1982gs}. Observables that are not invariant under $Z_N$ transformation are order parameters of this phase transition, and the Polyakov loop is a conventional choice among them~\cite{Polyakov:1978vu}.

When BCs along spatial directions are imposed into YM theory at nonzero $T$, the action has additional $Z_N$ symmetries corresponding to the individual BCs. In such systems, therefore, phase transitions associated with the spontaneous breaking of individual $Z_N$ symmetries are expected to occur with variations of $T$ and spatial extent along the BCs. These phenomena, which are regarded as generalized Casimir effects in thermal systems, have been investigated from various motivations~\cite{Simic:2010sv,Tiburzi:2013vza,Flachi:2013bc,Fraga:2016oul,Karabali:2018ael,Mogliacci:2018oea,Ishikawa:2018yey,Santos:2019xlx,Ishikawa:2019dcn,Ishikawa:2020ezm,Ishikawa:2020icy,Dudal:2020yah,Inagaki:2021yhi,Guo:2022mdi} including numerical simulations~\cite{Chernodub:2012em,Chernodub:2016owp,Chernodub:2017mhi,Chernodub:2017gwe,Chernodub:2018pmt,Chernodub:2018aix,Kitazawa:2019otp,Chernodub:2019kon,Chernodub:2022izt}.

Recently, thermodynamics of $SU(3)$ YM theory with a periodic boundary condition (PBC) along one spatial direction has been investigated in the lattice numerical simulation~\cite{Kitazawa:2019otp} using a technique based on the gradient flow~\cite{Suzuki:2013gza,Asakawa:2013laa,Kitazawa:2016dsl,Iritani:2018idk}. With the BC, the pressure becomes anisotropic due to the violation of the rotational symmetry~\cite{Brown:1969na}. The numerical results in Ref.~\cite{Kitazawa:2019otp}, however, show that in comparison with the free field theory the pressure anisotropy is remarkably suppressed until the spatial extent becomes significantly small near the critical temperature $T_c$. This result implies that the non-perturbative nature of the gauge field plays a crucial role in determining the response against the BC near $T_c$. It thus is also expected that the response can be used as a sensitive probe to understand the properties of the system.

In the present study, we investigate the phase transitions and thermodynamics of YM theory at nonzero $T$ with BCs using an effective model. To be specific, throughout this paper we take the spatial dimension to be three and impose the PBC for one direction, say $x$, with the length $L_x$. Since the BC for the temporal direction is also periodic for gauge fields, this system is defined on a Euclidean manifold $\mathbb{T}^2\times \mathbb{R}^2$ with two PBCs of length $L_\tau=1/T$ and $L_x$. This system has two $Z_N$ symmetries corresponding to two PBCs, and the phase transitions associated with their spontaneous breakings are expected to occur by varying $L_\tau$ and $L_x$. To describe these phase transitions simultaneously in an effective-model approach, the model must contain two order parameters.

For the YM theory on an infinite spatial volume without BCs, i.e. the theory on $\mathbb{S}^1\times \mathbb{R}^3$, an effective model including the Polyakov loop as the order parameter has been proposed in Ref.~\cite{Meisinger:2001cq}. In this model, the gauge field has a constant background field corresponding to the non-trivial expectation value of the Polyakov loop that is determined by the minimization of the free energy. It has been shown that thermodynamic quantities obtained on the lattice are qualitatively reproduced in the model. This result indicates that the Polyakov loop plays an important role in describing thermal properties of YM theory near $T_c$~\cite{Pisarski:2000eq}. The idea has been improved in the literature by incorporating various effects, and also applied to QCD with dynamical fermions; see Ref.~\cite{Fukushima:2017csk} as a review and references therein.

In the present study, to describe the theory on $\mathbb{T}^2\times \mathbb{R}^2$ we extend these models by introducing two ``Polyakov loops'' along two compactified directions. Using this model, we investigate their roles in thermal properties of the system, especially the emergence of the anisotropic pressure. As the first trial of such an attempt, we consider a simple model composed of massless gauge field on the background field and a simple trial form for the Polyakov-loop potential. We investigate the phase diagram on the $L_\tau$--$L_x$ plane and thermodynamics for $N=2$ and $3$.

We show that an interesting phase structure with the second- and first-order phase-transition lines emerges for $N=2$ and $3$, respectively. We also calculate thermodynamic quantities on $\mathbb{T}^2\times \mathbb{R}^2$ and compare them with the lattice results in Ref.~\cite{Kitazawa:2019otp}. In this analysis, we find that the thermodynamic quantities are significantly affected by the two Polyakov loops. We, however, find that our model fails to reproduce the lattice results even qualitatively. While the result obviously shows that our approach needs to be improved, we argue that such a modification is possible by changing the form of the potential term. Such a description of the YM theory on $\mathbb{T}^2\times \mathbb{R}^2$ will in turn give us a deeper understanding of the theory on $\mathbb{S}^1\times \mathbb{R}^3$. Although we do not attempt further modifications of the model in this exploratory study, possible directions will be discussed.

This article is organized as follows. In Sec.~\ref{sec:model}, we introduce the effective model for the YM theory on $\mathbb{T}^2\times \mathbb{R}^2$ including two types of the Polyakov loops. Besides, we explain its properties and present concrete expressions for $N=2$ and $N=3$. In Sec.~\ref{sec:PhaseDiagram}, we show our numerical results of the phase diagram on the $L_\tau$--$L_x$ plane for $N=2$ and $N=3$. In Sec.~\ref{sec:Thermodynamics} our model is compared to the lattice data of thermodynamic quantities obtained in Ref.~\cite{Kitazawa:2019otp} and some discussions on the role of the Polyakov loops are provided. Then, our present work is concluded in Sec.~\ref{sec:Conclusions}.

\section{Model}
\label{sec:model}

In the present study we introduce an effective model for describing the YM theory on the $\mathbb{T}^2\times \mathbb{R}^2$ Euclidean manifold with PBCs. This system physically corresponds to a thermal system with the PBC along one spatial direction additionally, where the temporal length is related to temperature as $L_\tau=1/T$. We suppose that the PBC is imposed along $x$ direction and define the spatial extent by $L_x$, while the lengths along $y$ and $z$ directions are taken to be infinite.

\subsection{Polyakov loops}

The YM theory on $\mathbb{T}^2\times \mathbb{R}^2$ is invariant under the center symmetries, which are denoted as $Z_N^{(\tau)}$ and $Z_N^{(x)}$. These symmetries can be spontaneously broken with variations of $L_\tau$ and $L_x$. To describe these phase transitions, we employ the Polyakov loops\footnote{
  The quantity that is conventionally called the Polyakov loop is $P_\tau$. In this manuscript, however, we refer to both $P_\tau$ and $P_x$ as the Polyakov loops for simplicity.
}
\begin{eqnarray}
P_\tau(x,{\bm r}_L) &=& \frac1N {\rm Tr} \left[ \hat{P}_\tau(x,{\bm r}_L)   \right]\ , \nonumber\\
P_x(\tau,{\bm r}_L) &=& \frac1N {\rm Tr} \left[\hat{P}_x(\tau,{\bm r}_L)  \right] \ , \label{PDefinition}
\end{eqnarray}
for the order parameters, where Tr means the trace over the gauge space and the Polyakov-loop matrices are defined by
\begin{eqnarray}
\hat{P}_\tau (x,{\bm r}_L) &=& \mathscr{P}\, {\rm exp}\left(i\int_0^{L_\tau}{A}_\tau(\tau,x,{\bm r}_L) d\tau \right)\ , \nonumber\\
\hat{P}_x (\tau,{\bm r}_L) &=&  \mathscr{P}\, {\rm exp}\left(i\int_0^{L_x}{A}_x(\tau,x,{\bm r}_L) dx \right)\ . \label{PHatDef}
\end{eqnarray}
In Eq.~(\ref{PHatDef}), $A_\mu(\tau,x,\bm{r}_L)$ is the YM gauge field [$\mu=(x,y,z,\tau)$], $\mathscr{P}$ stands for the path-ordering integral, and ${\bm r}_L=(y,z)$ represents the uncompactified two spatial variables. 

The Polyakov loops $P_\tau$ and $P_x$ are not invariant under $Z_N^{(\tau)}$ and $Z_N^{(x)}$, respectively, and are order parameters of the corresponding symmetries. It is also known that the thermal expectation value $\langle P_\tau \rangle$ is related to the free energy $F_q(x,{\bm r}_L)$ of a static test quark as~\cite{Rothe:1992nt}
\begin{eqnarray}
{\rm e}^{-L_\tau F_q(x,{\bm r}_L)} = N \langle P_\tau(x,{\bm r}_L) \rangle \ .
\end{eqnarray}
Therefore, $\langle P_\tau \rangle=0$ corresponds to the confined phase in which $F_q$ becomes infinite. In YM theory on $\mathbb{S}^1\times\mathbb{R}^3$, this phase is realized as a low-temperature phase, while the deconfined phase with $\langle P_\tau \rangle\ne0$ is realized at high $T$. These phases are separated by the second- and first-order phase transition for $N=2$ and $N\ge3$, respectively~\cite{Svetitsky:1982gs,Yaffe:1982qf,Svetitsky:1985ye,Sannino:2002wb,Ruggieri:2012ny,Sasaki:2012bi,Sasaki:2013xfa,Fukushima:2017csk}.

\subsection{Model construction}

For a simultaneous description of the spontaneous breakings of $Z_N^{(\tau)}$ and $Z_N^{(x)}$ on $\mathbb{T}^2\times \mathbb{R}^2$, we construct an effective model including $P_\tau$ and $P_x$ for their order parameters. To this end, we extend the ``model-B'' introduced in Ref.~\cite{Meisinger:2001cq} that deals with a single Polyakov loop $P_\tau$ on $\mathbb{S}^1\times \mathbb{R}^3$.

In Ref.~\cite{Meisinger:2001cq}, to represent a non-trivial value of $P_\tau$ the ``mean-field'' assumption is imposed, where the gauge field has a constant background field. The free energy of the system consists of two parts; the first one is the one-loop perturbative contribution from the gauge field with the background field, and the other is a potential term that models non-perturbative effects leading to the confinement. The expectation value of $P_\tau$ is determined so as to minimize the free energy. It has been shown that the model can reproduce $T$ dependence of thermodynamic quantities measured in lattice simulations qualitatively.

By extending this idea, we assume that the gauge field on $\mathbb{T}^2\times \mathbb{R}^2$ has a constant background field
\begin{eqnarray}
  A_\tau(\tau,x,\bm{r}_L) &=& \frac{1}{L_\tau} {\rm diag}\Big[(\theta_\tau)_1, (\theta_\tau)_2, \cdots, (\theta_\tau)_{N} \Big]\ ,
  \label{ABGAnsatz_t}\\
A_x(\tau,x,\bm{r}_L) &=& \frac{1}{L_x} {\rm diag}\Big[(\theta_x)_1, (\theta_x)_2, \cdots, (\theta_x)_{N} \Big]\ ,  \label{ABGAnsatz_x}
\end{eqnarray}
corresponding to the expectation values of $P_\tau$ and $P_x$.
Substituting Eqs.~(\ref{ABGAnsatz_t}) and~(\ref{ABGAnsatz_x}) into Eq.~(\ref{PHatDef}), one obtains
\begin{eqnarray}
  (\hat{P}_c)_{jk} = \exp( i(\theta_c)_j ) \delta_{jk} \ ,
\end{eqnarray}
for $c=\tau,~x$ and 
\begin{eqnarray}
  P_c = \frac1N \sum_{j=1}^N \exp( i(\theta_c)_j ) \ .
\end{eqnarray}
Since $\hat{P}_c$ is an element of $SU(N)$, the phase variables $\vec\theta_c=( (\theta_c)_1, \cdots, (\theta_c)_{N} )$ satisfy
\begin{eqnarray}
  \sum_{j=1}^N (\theta_c)_j = 0 \quad \mbox{(mod $2\pi$)}\ .
  \label{theta=0}
\end{eqnarray}

In Eqs.~(\ref{ABGAnsatz_t}) and~(\ref{ABGAnsatz_x}), the background field is assumed to have diagonal forms following Ref.~\cite{Meisinger:2001cq}. In Ref.~\cite{Meisinger:2001cq}, the background field is introduced only for $A_\tau(\tau,x,\bm{r}_L)$ because $P_x$ is irrelevant on $\mathbb{S}^1\times \mathbb{R}^3$. In this case, $A_\tau(\tau,x,\bm{r}_L)$ can be diagonalized only with the gauge transformation. In contrast, the simultaneous diagonalization of $A_\tau(\tau,x,\bm{r}_L)$ and $A_x(\tau,x,\bm{r}_L)$ as in Eqs.~(\ref{ABGAnsatz_t}) and~(\ref{ABGAnsatz_x}) by using gauge degrees of freedom is not possible in general. However, since the purpose of the present study is to explore qualitative effects of $P_\tau$ and $P_x$ on $\mathbb{T}^2\times \mathbb{R}^2$, we employ the ansatz~(\ref{ABGAnsatz_t}) and (\ref{ABGAnsatz_x}), which is a simple choice to realize non-trivial values of $P_\tau$ and $P_x$.

For later use, it is convenient to parametrize the phase variables $\vec\theta_\tau$ and $\vec\theta_x$ for $N=2$ and $N=3$ within the constraint~(\ref{theta=0}). For $N=2$, each phase is parametrized by a single parameter $\phi_c$ as
\begin{eqnarray}
  \vec\theta_c = ( \phi_c , -\phi_c ) \ .
  \label{AnsatzSU2}
\end{eqnarray}
In this parametrization the Polyakov loops are given by
\begin{eqnarray}
P_c = \cos\phi_c \ .
\end{eqnarray}
The non-perturbative vacuum where the center symmetries are restored is given by $\phi_c=\pi/2$. On the other hand, the completely perturbative vacuum is realized with $\phi_c=0$. We thus impose the range of $\phi_c$ as $0\leq \phi_c \leq \pi/2$ for $N=2$.

For $N=3$, $\vec{\theta}_c$ has two degrees of freedom. The symmetric phase with $P_c=0$ is realized, for example, at $\vec\theta_c=(2\pi/3,0,-2\pi/3)$, while the perturbative vacuum corresponds to $\vec\theta_c=\vec0$. To connect these values, following Ref.~\cite{Meisinger:2001cq,Fukushima:2017csk}, we parametrize the phases as
\begin{eqnarray}
  \vec\theta_c = (\phi_c, 0, -\phi_c ) \ ,
  \label{AnsatzSU3}
\end{eqnarray}
resulting in 
\begin{eqnarray}
  P_c = \frac{1}{3}( 1+2\cos\phi_c)  \ ,
\end{eqnarray}
with $0\leq\phi_c\leq2\pi/3$. 

The one-loop free energy per unit volume with the background field~(\ref{ABGAnsatz_t}) and~(\ref{ABGAnsatz_x}) is calculated to be~\cite{Sasaki:2012bi}
\begin{eqnarray}
  && f_{\rm pert}( \vec\theta_\tau,\vec\theta_x ; L_\tau,L_x )
  \nonumber \\
  &&= \frac{2}{L_\tau L_x}\sum_{j,k=1}^{N}\left(1-\frac{\delta_{jk}}{N}\right)\sum_{l_\tau,l_x}\int\frac{d^2p_L}{(2\pi)^2} \nonumber\\
  && \times {\rm ln}\left[\left(\omega_\tau-\frac{(\Delta\theta_\tau)_{jk}}{L_\tau}\right)^2 + \left(\omega_x+\frac{(\Delta\theta_x)_{jk}}{L_x}\right)^2 + {\bm p}^2_L\right]\ , \nonumber\\
  \label{FPert}
\end{eqnarray}
where the mass of the gauge field is assumed to vanish. In Eq.~(\ref{FPert}), $\omega_\tau=2\pi l_\tau/L_\tau$ and $\omega_x=2\pi l_x/L_x$ are the respective ``Matsubara modes'' with $l_\tau$ and $l_x$ being integers. We have defined momenta for the uncompactified directions by ${\bm p}_L=(p_y,p_z)$, and the phase differences $(\Delta\theta_c)_{jk}=(\theta_c)_j-(\theta_c)_k$.

For a given set of $(L_\tau,L_x)$, the minimum of Eq.~(\ref{FPert}) is at $\vec\theta_\tau=\vec\theta_x=\vec0$ (mod $2\pi$), which gives $P_\tau=P_x=1$. Defining the expectation values of $P_\tau$ and $P_x$ as the minimum of Eq.~(\ref{FPert}), therefore, the system is always in the deconfined phase.  To model the confinement phase transition, we introduce a potential term of the Polyakov loops $f_{\rm pot}( \vec\theta_\tau,\vec\theta_x; L_\tau , L_x )$, such that the total free energy per unit volume reads
\begin{align}
  f( \vec\theta_\tau,\vec\theta_x; L_\tau , L_x )
  =& f_{\rm pert}( \vec\theta_\tau,\vec\theta_x; L_\tau , L_x )
  \nonumber \\
  &+ f_{\rm pot}( \vec\theta_\tau,\vec\theta_x; L_\tau , L_x ) \ .
  \label{F}
\end{align}
Here, $f_{\rm pot}$ describes the non-perturbative effects that would have a dominant contribution at large $L_\tau$ or $L_x$. The specific form of $f_{\rm pot}$ will be determined later. The values of $\vec\theta_c$, and hence $P_c$,  are determined so as to minimize Eq.~(\ref{F}).

\subsection{One-loop free energy}

With the parametrizations~(\ref{AnsatzSU2}) and (\ref{AnsatzSU3}), the one-loop free energy~(\ref{FPert}) is transformed into the form that is more suitable for numerical analyses. After the manipulations summarized in Appendix~\ref{sec:FPert} with the regularization explained in Appendix~\ref{sec:ZetaFunction}, one obtains
\begin{eqnarray}
f_{\rm pert} &=& - \frac{\pi^2}{15L_\tau^4} + \frac{4\phi_\tau^2(\phi_\tau-\pi)^2}{3\pi^2L_\tau^4}  -\frac{\pi^2}{15L_x^4} + \frac{4\phi_x^2(\phi_x-\pi)^2 }{3\pi^2L_x^4} \nonumber\\
&&-\frac{4}{\pi^{2}}\sum_{l_\tau,l_x=1}^\infty\frac{1+2\cos(2\phi_\tau l_\tau)\cos(2\phi_xl_x)}{X_{l_\tau,l_x}^4} \ , \label{FPertSU2}
\end{eqnarray}
for $N=2$, while for $N=3$ one has
\begin{eqnarray}
f_{\rm pert} &=&  -\frac{8\pi^2}{45L_\tau^4}+ \frac{8\phi_\tau^2(\phi_\tau-\pi)^2+\phi_\tau^2(\phi_\tau-2\pi)^2}{6\pi^2L_\tau^4} \nonumber\\
&&- \frac{8\pi^2}{45L_x^4} + \frac{8\phi_x^2(\phi_x-\pi)^2+\phi_x^2(\phi_x-2\pi)^2}{6\pi^2L_x^4} \nonumber\\
&&-\frac{8}{\pi^{2}}\sum_{l_\tau,l_x=1}^\infty\frac{1}{X_{l_\tau,l_x}^4} \Big[1+ 2\cos(\phi_\tau l_\tau)\cos(\phi_xl_x) \nonumber\\
&&+ \cos(2\phi_\tau l_\tau) \cos(2\phi_xl_x)\Big] \ , \label{FPertSU3}
\end{eqnarray}
with
\begin{eqnarray}
  X_{l_\tau,l_x} \equiv \sqrt{(l_\tau L_\tau )^2+(l_xL_x)^2} \ .
  \label{X}
\end{eqnarray}

In the limit $L_x\to\infty$, only the first two terms in Eqs.~(\ref{FPertSU2}) and (\ref{FPertSU3}) that do not include $1/L_x$ survive. The form of $f_{\rm pert}$ in this limit reproduces that for $\mathbb{S}^1\times\mathbb{R}^3$~\cite{Meisinger:2001cq}.
The subleading terms with respect to $1/L_x$ come from the last double-sum term in Eq.~(\ref{FPertSU2}) or (\ref{FPertSU3}). As discussed in Appendix~\ref{sec:FPert}, the subleading term starts at the order $(L_\tau/L_x)^3$;
\begin{eqnarray}
  f_{\rm pert} &=& \hat{f}_{\infty}(\vec\theta_\tau) \frac1{L_\tau^4} 
  + \mathcal{O}(L_x^{-3}) \ .
  \label{Finf_exp}
\end{eqnarray}
Note that odd-order terms of $L_\tau/L_x$ appear in this expansion, while Eqs.~(\ref{FPertSU2}) and (\ref{FPertSU3}) apparently depend on $L_\tau/L_x$ only through $(L_\tau/L_x)^2$; see Appendix~\ref{sec:FPert}.

\subsection{Potential term}

Next, we determine the form of $f_{\rm pot}$. To this end let us first consider general properties of this term.

First, since the YM theory on $\mathbb{T}^2\times \mathbb{R}^2$ is invariant under the exchange of the $\tau$ and $x$ axes, $f_{\rm pot}$ should satisify
\begin{eqnarray}
  f_{\rm pot}( \vec\theta_\tau,\vec\theta_x ; L_\tau , L_x )
  = f_{\rm pot}( \vec\theta_x,\vec\theta_\tau ; L_x , L_\tau ) \ .
  \label{Fsym}
\end{eqnarray}

Second, in the $L_\tau\to\infty$ ($T\to0$) limit the system with the PBC would be in the confined phase irrespective of the value of $L_x$. We thus have
\begin{eqnarray}
  P_\tau = 0 \quad (L_\tau\to\infty) \ .
  \label{PtLt->inf}
\end{eqnarray}
By exchanging the $\tau$ and $x$ axes, one also obtains
\begin{eqnarray}
  P_x = 0 \quad (L_x\to\infty) \ .
  \label{PxLx->inf}
\end{eqnarray}
It is worth emphasizing that the limiting value of $P_c$ in these limits is not the perturbative one $P_c=1$. However, the values of $P_\tau$ and $P_x$ in these limits are irrelevant in the sense that they do not affect the property of the system since effects of the BC should be negligible. In fact, the background field $A_c$ in Eqs.~(\ref{ABGAnsatz_t}) and~(\ref{ABGAnsatz_x}) vanishes in the $L_c\to\infty$ limit irrespective of the value of $\vec\theta_c$. 

Third, since the system approaches $\mathbb{S}^1\times \mathbb{R}^3$ in the limit $L_x\to\infty$, the potential term should approach the one on $\mathbb{S}^1\times \mathbb{R}^3$, i.e.
\begin{eqnarray}
  f_{\rm pot}( \vec\theta_\tau,\vec\theta_x ; L_\tau , L_x )
  \underset{L_x\to\infty}\longrightarrow f_{\rm pot}^{\mathbb{S}^1\times\mathbb{R}^3}( \vec\theta_\tau ; L_\tau ) \ ,
  \label{Fpot_lim1}
\end{eqnarray}
where $f_{\rm pot}^{\mathbb{S}^1\times\mathbb{R}^3}( \vec\theta ; L )$ is the potential term for the effective model for $\mathbb{S}^1\times \mathbb{R}^3$ with a single Polyakov loop $P_\tau$.
Equation~(\ref{Fpot_lim1}) means that $\vec\theta_x$ dependence in $f_{\rm pot}$ exists at the subleading order of $L_\tau/L_x$,
and this contribution leads to Eq.~(\ref{PxLx->inf}). This implies that the $\vec\theta_x$ dependence in $f_{\rm pot}$ must surpass the perturbative contributions $f_{\rm pert}$ which is of order $(L_\tau/L_x)^3$ as in Eq.~(\ref{Finf_exp}), and hence the former should be weaker than the power of $(L_\tau/L_x)^3$. From Eq.~(\ref{Fsym}), one obtains the same conclusion for the limit $L_\tau\to\infty$, i.e.
\begin{eqnarray}
  f_{\rm pot}( \vec\theta_\tau,\vec\theta_x ; L_\tau , L_x )
  \underset{L_\tau\to\infty}\longrightarrow f_{\rm pot}^{\mathbb{S}^1\times\mathbb{R}^3}( \vec\theta_x ; L_x ) \ ,
  \label{Fpot_lim2}
\end{eqnarray}
with $\vec\theta_\tau$ dependence satisfying Eq.~(\ref{PtLt->inf}).
When one constructs an effective model on $\mathbb{T}^2\times \mathbb{R}^2$ as an extension of that developed for $\mathbb{S}^1\times \mathbb{R}^3$, Eqs.~(\ref{Fpot_lim1}) and (\ref{Fpot_lim2}) are constraints for $f_{\rm pot}$.

In the model-B of Ref.~\cite{Meisinger:2001cq}, as an effective model on $\mathbb{S}^1\times \mathbb{R}^3$ the form of the potential term motivated by the Haar measure in the strong-coupling expansion
\begin{eqnarray}
  f_{\rm pot}^{\mathbb{S}^1\times\mathbb{R}^3}( \vec\theta_\tau ; L )
  = -\frac{1}{L R^3}{\rm ln}\bigg[\prod_{j<k}\sin^2\left(\frac{(\Delta \theta_\tau)_{jk}}{2}\right)\bigg] \ ,
  \label{FHaarinf}
\end{eqnarray}
has been employed in combination with $f_{\rm pert}$ for the massless gauge field. Here, the quantity $R$ having a mass dimension of $-1$ is understood as a typical length scale for the confinement; when $R\ll L_\tau$, the potential term dominates over $f_{\rm pert}$ and the confined phase is realized. 

In the present study, we employ Eq.~(\ref{FHaarinf}) as the form of $f_{\rm pot}^{\mathbb{S}^1\times\mathbb{R}^3}$ and introduce $f_{\rm pot}( \vec\theta_\tau,\vec\theta_x ; L_\tau , L_x )$ so as to satisfy Eqs.~(\ref{Fpot_lim1}) and (\ref{Fpot_lim2}). There are, of course, infinitely many possible forms of $f_{\rm pot}( \vec\theta_\tau,\vec\theta_x ; L_\tau , L_x )$ within the constraint. Among them, in this exploratory study, we employ a simple seprable ansatz 
\begin{eqnarray}
  \lefteqn{ f_{\rm pot}( \vec\theta_\tau,\vec\theta_x ; L_\tau,L_x ) }
  \nonumber \\
  &=&
  f_{\rm pot}^{\mathbb{S}^1\times\mathbb{R}^3}( \vec\theta_\tau ; L_\tau ) + f_{\rm pot}^{\mathbb{S}^1\times\mathbb{R}^3}( \vec\theta_x ; L_x ) \nonumber \\
  &=&-\frac{1}{L_\tau R^3}{\rm ln}\bigg[\prod_{j<k}\sin^2\left(\frac{(\Delta \theta_\tau)_{jk}}{2}\right)\bigg] \nonumber\\
&&- \frac{1}{L_x R^3}{\rm ln}\bigg[\prod_{j<k}\sin^2\left(\frac{(\Delta \theta_x)_{jk}}{2}\right)\bigg]\ . \label{FHaar}
\end{eqnarray}
We note that Eq.~(\ref{FHaar}) satisfies all the conditions~(\ref{Fsym})--(\ref{Fpot_lim2}). 

With the phase variables in Eqs.~(\ref{AnsatzSU2}) and~(\ref{AnsatzSU3}), Eq.~(\ref{FHaar}) is reduced to 
\begin{eqnarray}
f_{\rm pot} &=& -\frac{1}{L_\tau R^3}{\rm ln}(\sin^2\phi_\tau)-\frac{1}{L_x R^3}{\rm ln}(\sin^2\phi_x) \ ,\nonumber\\
\label{FHaarSU2}
\end{eqnarray}
for $N=2$ and 
\begin{eqnarray}
f_{\rm pot} &=& -\frac{1}{L_\tau R^3}{\rm ln}\left[\left(\sin^4\frac{\phi_\tau}{2}\right)\left(\sin^2\phi_\tau\right)\right]  \nonumber\\
&&-  \frac{1}{L_xR^3}{\rm ln}\left[\left(\sin^4\frac{\phi_x}{2}\right)\left(\sin^2\phi_x\right)\right]\ ,   \label{FHaarSU3}
\end{eqnarray}
for $N=3$.

\section{Phase diagram}
\label{sec:PhaseDiagram}

In this section, we investigate the model introduced in the previous section focusing on the behavior of the order parameters and the phase diagram on the $L_\tau$--$L_x$ plane for $N=2,3$.

\subsection{$N=2$}
\label{sec:ResultsSU2}

\begin{figure}[t]
\centering
\hspace*{-0.5cm} 
\includegraphics*[scale=0.58]{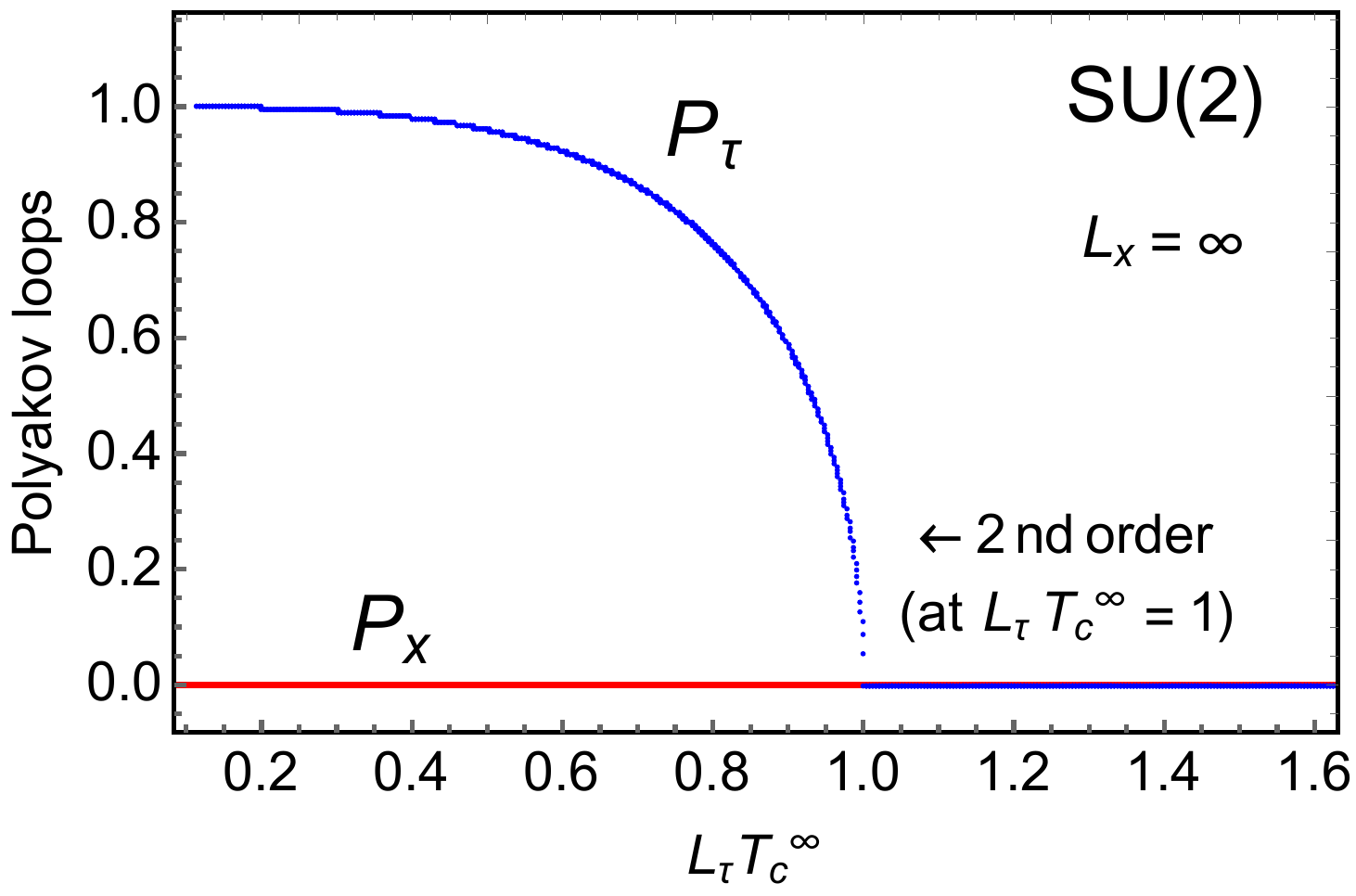}
\caption{$L_\tau$ dependence of $P_\tau$ (blue) and $P_x$ (red) at $L_x\to\infty$ for $N=2$. There exists a second-order phase transition at $L_\tau T_c^\infty=1$ with the critical temperature $T_c^\infty\approx 1/(0.874R)$. The spatial Polyakov loop $P_x$ is always zero.}
\label{fig:FiniteTSU2}
\end{figure}

\begin{figure}[hbtp]
  \begin{center}
    \begin{tabular}{cc}

      \begin{minipage}[c]{0.47\hsize}
       \centering
       \hspace*{-2.5cm} 
         \includegraphics*[scale=0.56]{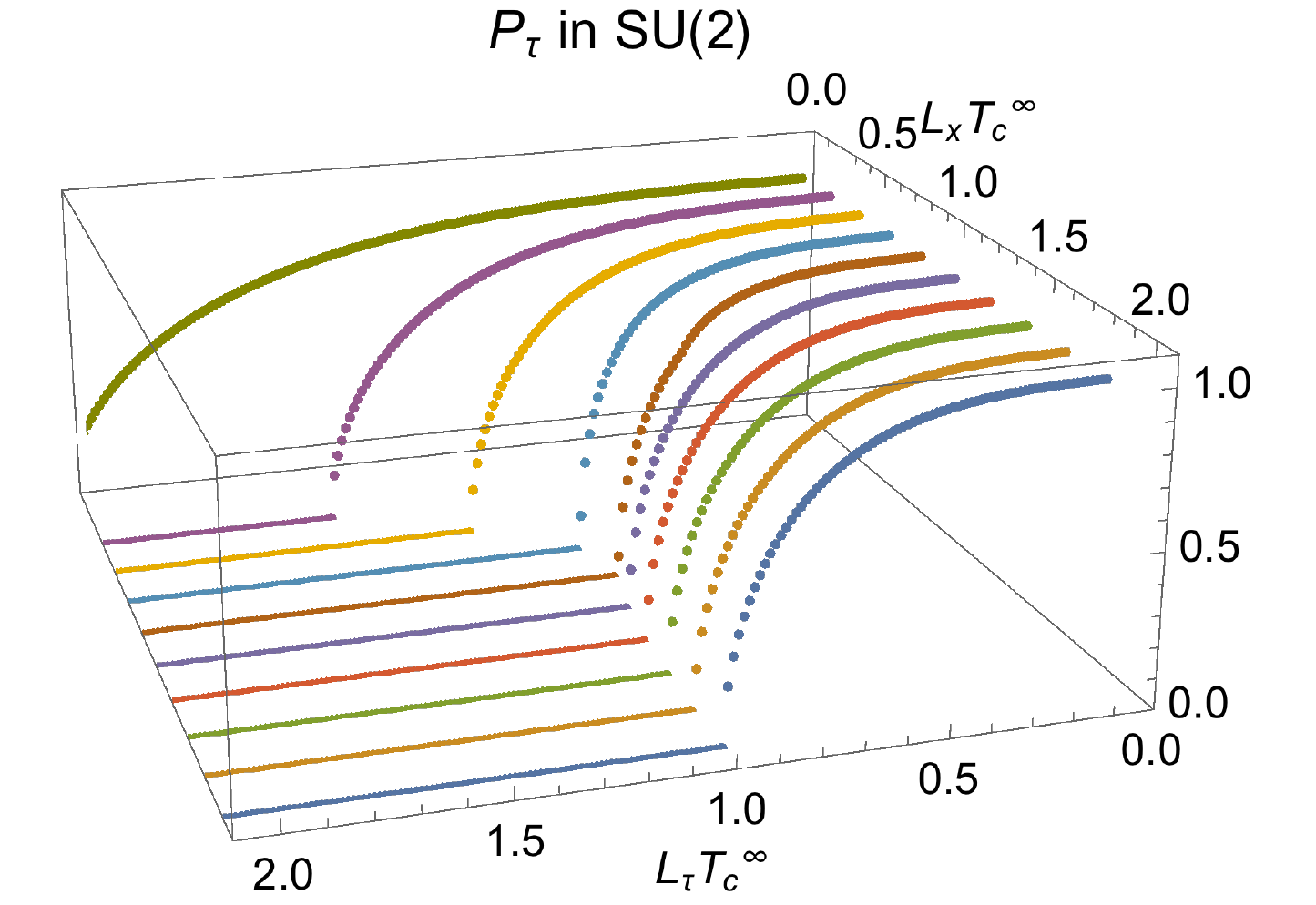}\\
         \end{minipage}
\\
\\
\\
      \begin{minipage}[c]{0.4\hsize}
       \centering
        \hspace*{-2.6cm} 
          \includegraphics*[scale=0.56]{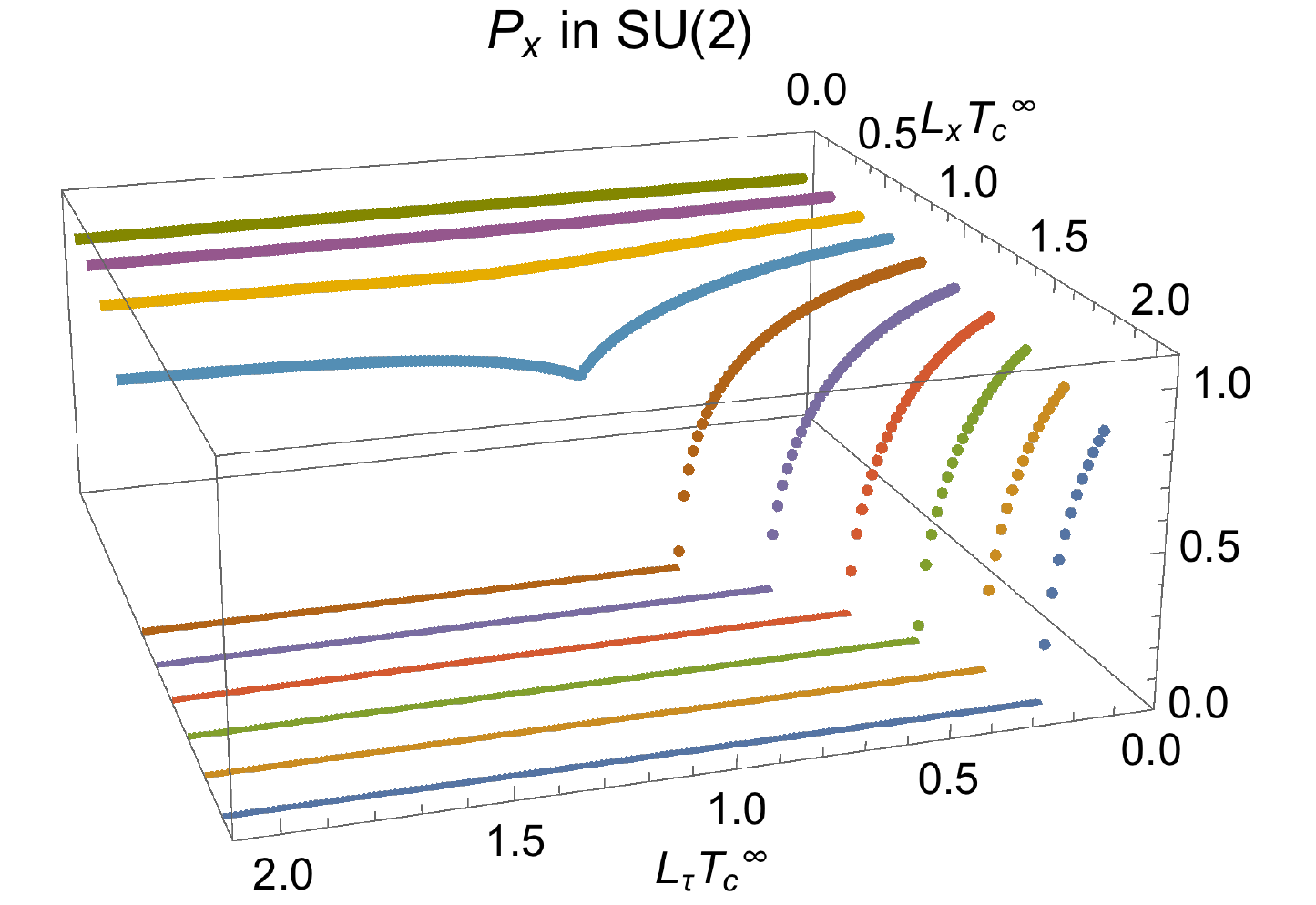}\\
      \end{minipage}

      \end{tabular}
 \caption{$L_\tau$ dependences of $P_\tau$ (top panel) and $P_x$ (bottom panel) at several $L_x$ for $N=2$.} 
\label{fig:3DPsiSU2}
  \end{center}
\end{figure}

\begin{figure}[t]
\centering
\hspace*{-0.5cm} 
\includegraphics*[scale=0.58]{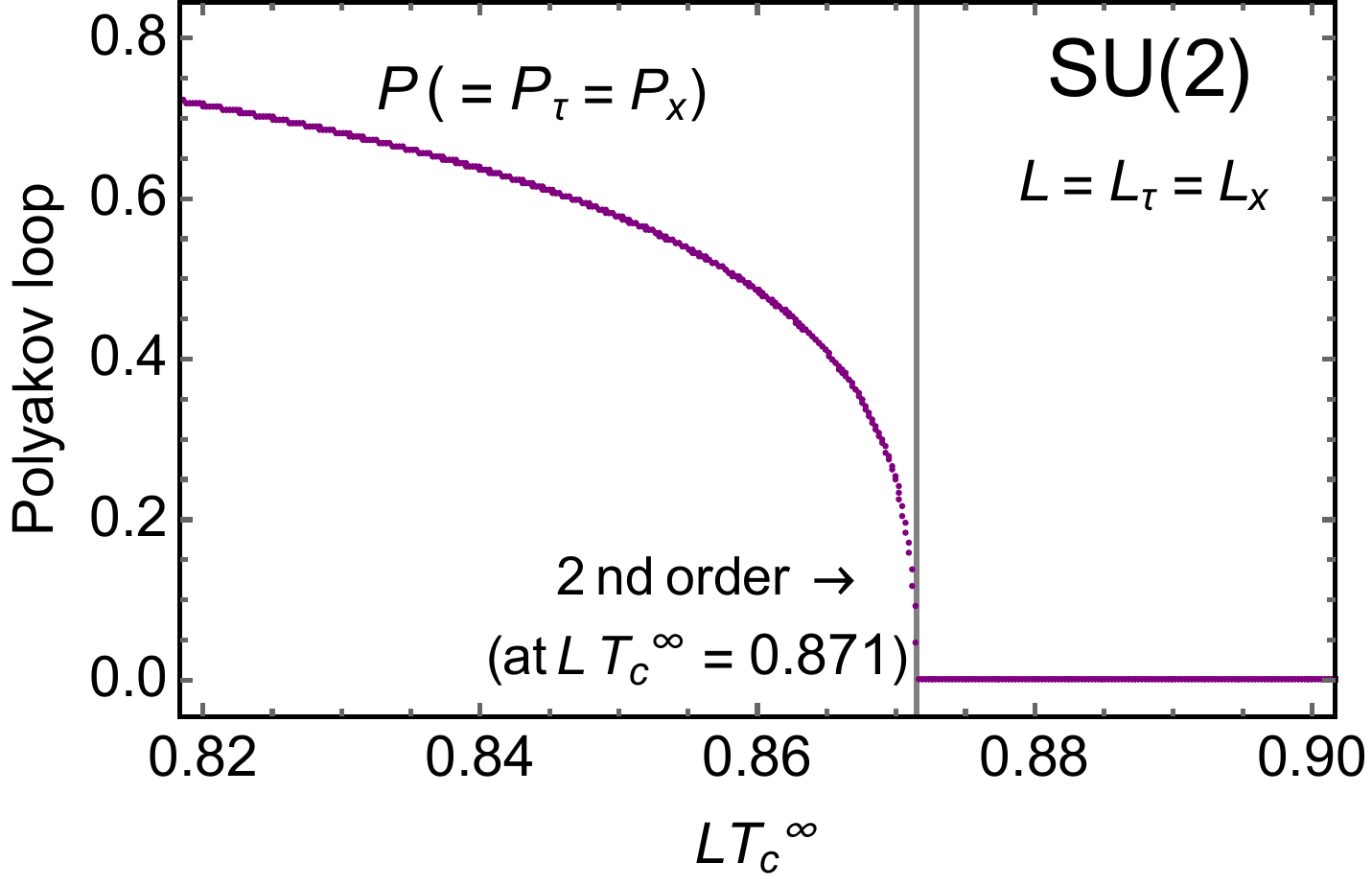}
\caption{Polyakov loops $ P(\equiv P_\tau=P_x)$ for the symmetric case $L=L_\tau=L_x$ for $N=2$ in the vicinity of the transition point. The second-order transition is found at $LT^\infty_c\approx 0.871$. The vertical line corresponds to the analytic solution for the second-order transition evaluated in Eq.~(\ref{AnalSU2}).}
\label{fig:SymSU2}
\end{figure}

\begin{figure}[t]
\centering
\hspace*{-0.5cm} 
\includegraphics*[scale=0.54]{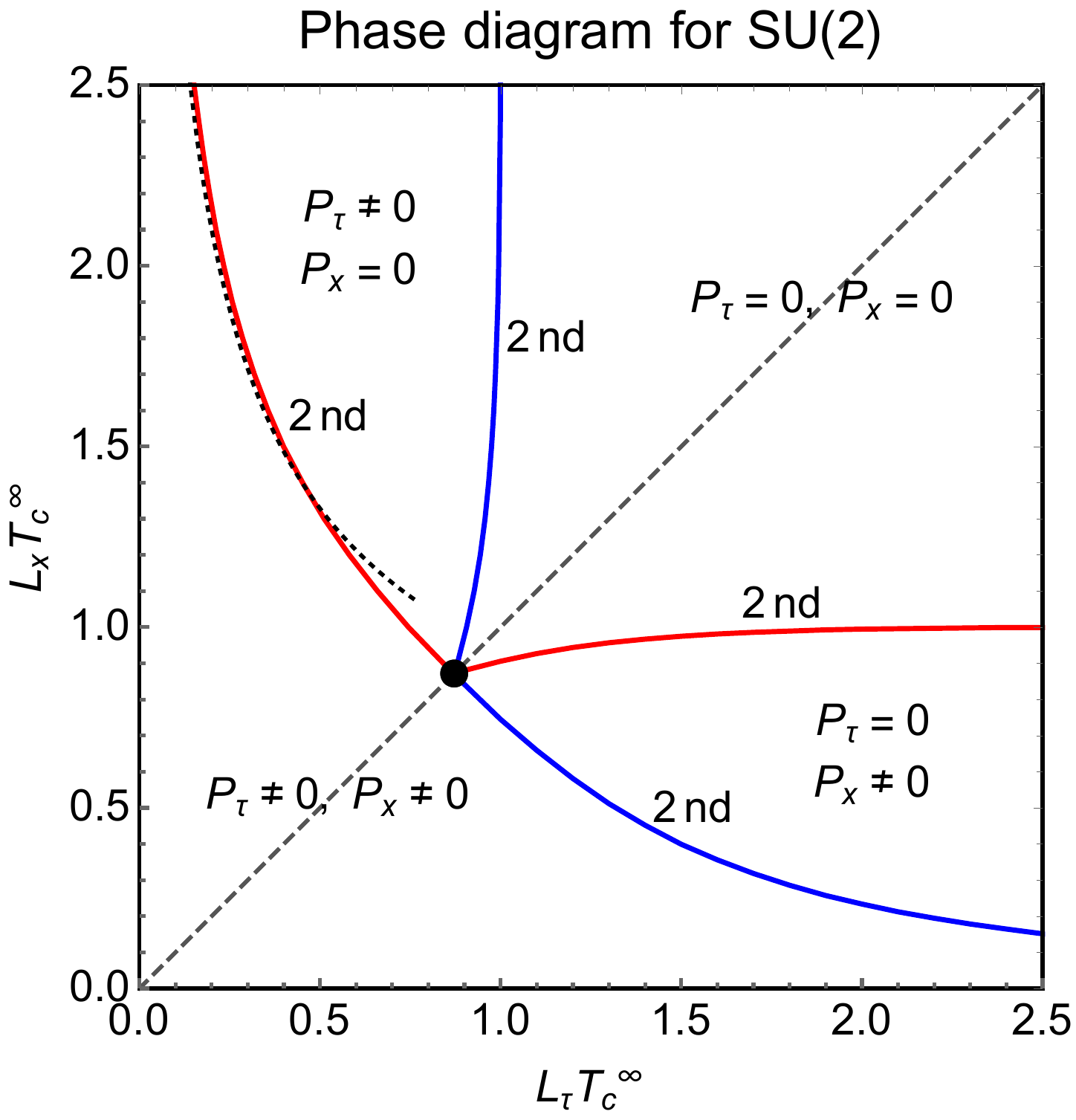}
\caption{Phase diagram on the $L_\tau$--$L_x$ plane for $N=2$. The blue (red) line separates $P_\tau=0$ and $P_\tau\neq0$ ($P_x=0$ and $P_x\neq0$). The dashed gray line stands for $L_\tau=L_x$, and the black dot is the transition point in Fig.~\ref{fig:SymSU2}. All phase transitions are of second order. The dotted line shows the asymptotic formula~(\ref{crit}).}
\label{fig:PhaseSU2}
\end{figure}

\begin{figure}[tb]
\centering
\hspace*{-0.5cm} 
\includegraphics*[scale=0.58]{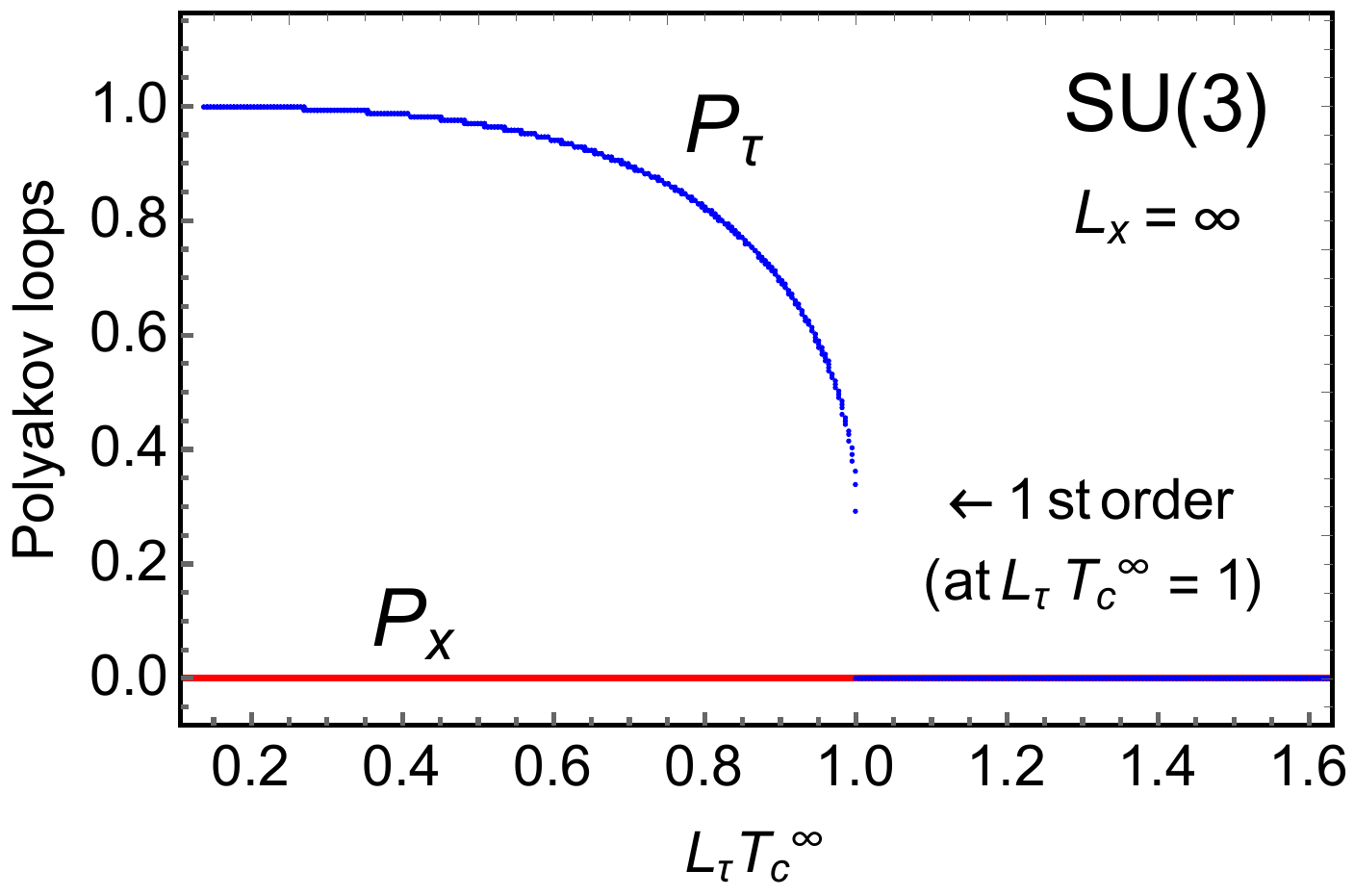}
\caption{$L_\tau$ dependence of $P_\tau$ (blue) and $P_x$ (red) at $L_x\to\infty$ for $N=3$. The temporal Polyakov loop $P_\tau$ shows a first-order phase transition at $T_c^\infty \approx 1/(0.733R)$.}
\label{fig:FiniteTSU3}
\end{figure}

\begin{figure}[hbtp]
  \begin{center}
    \begin{tabular}{cc}

      \begin{minipage}[c]{0.47\hsize}
       \centering
       \hspace*{-2cm} 
         \includegraphics*[scale=0.54]{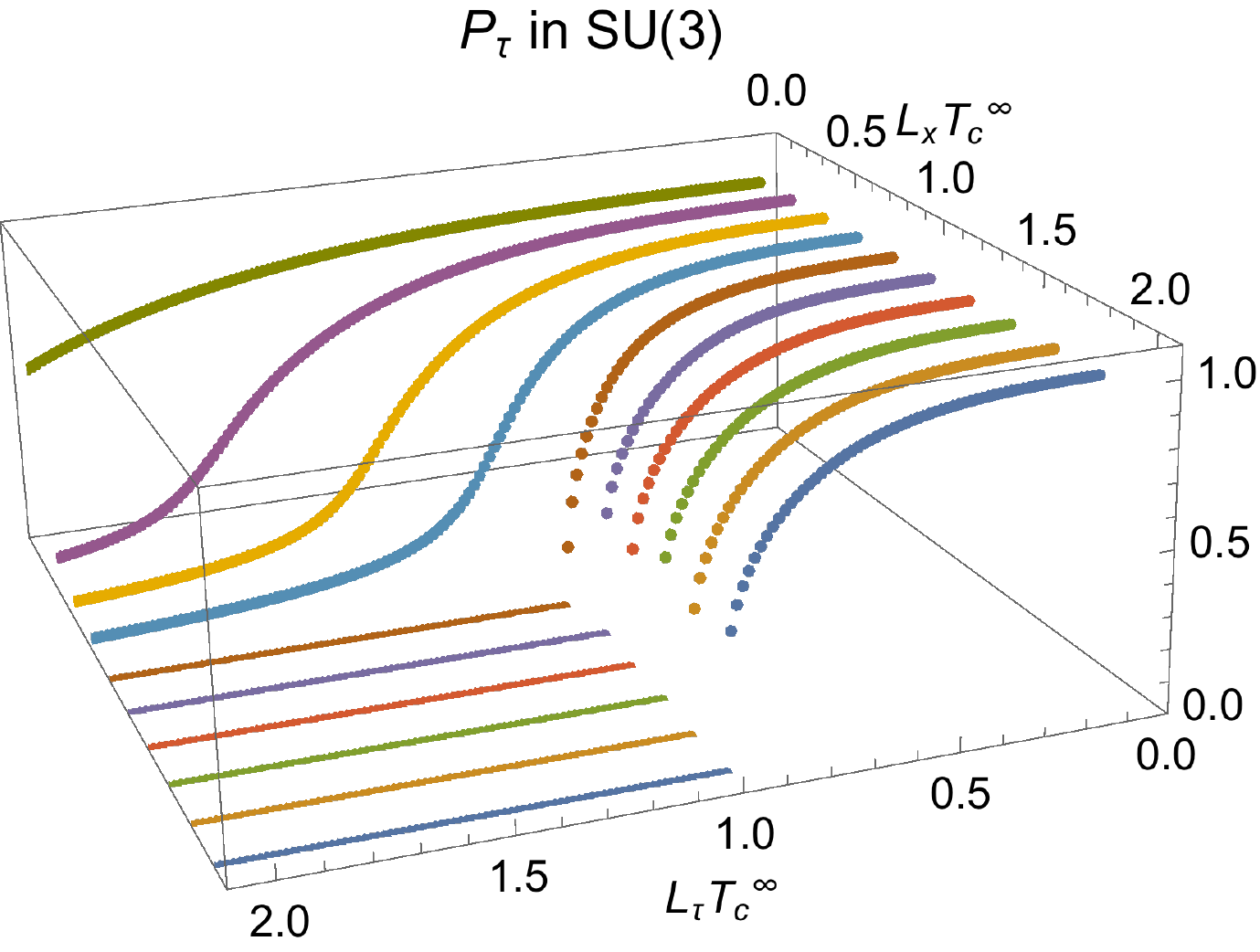}\\
         \end{minipage}
\\
\\
\\
      \begin{minipage}[c]{0.4\hsize}
       \centering
        \hspace*{-2.1cm} 
          \includegraphics*[scale=0.54]{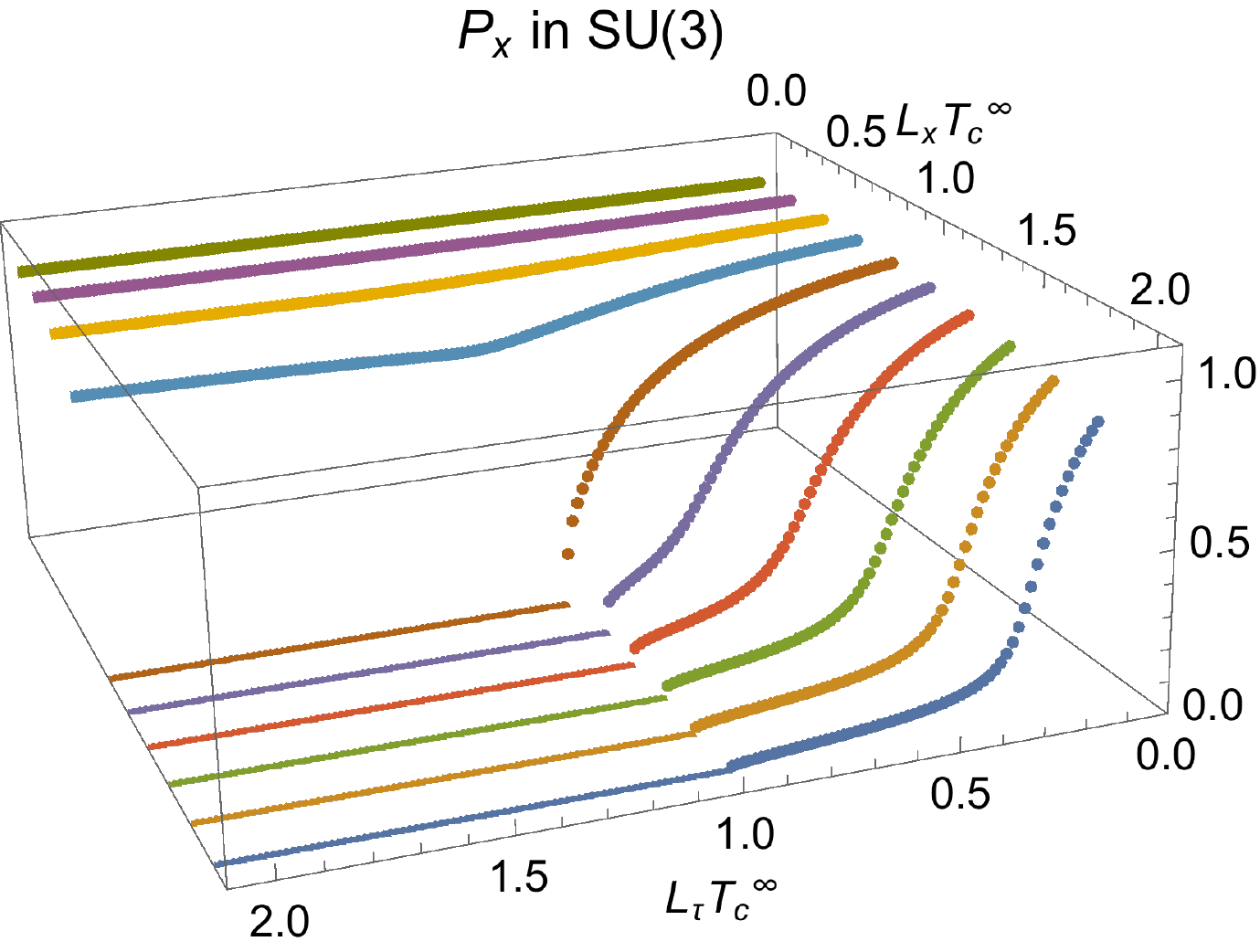}\\
      \end{minipage}

      \end{tabular}
 \caption{$L_\tau$ dependences of $P_\tau$ (top panel) and $P_x$ (bottom panel) at several $L_x$ for $N=3$.} 
\label{fig:3DPsiSU3}
  \end{center}
\end{figure}

\begin{figure}[hbtp]
\centering
\hspace*{-0.5cm} 
\includegraphics*[scale=0.58]{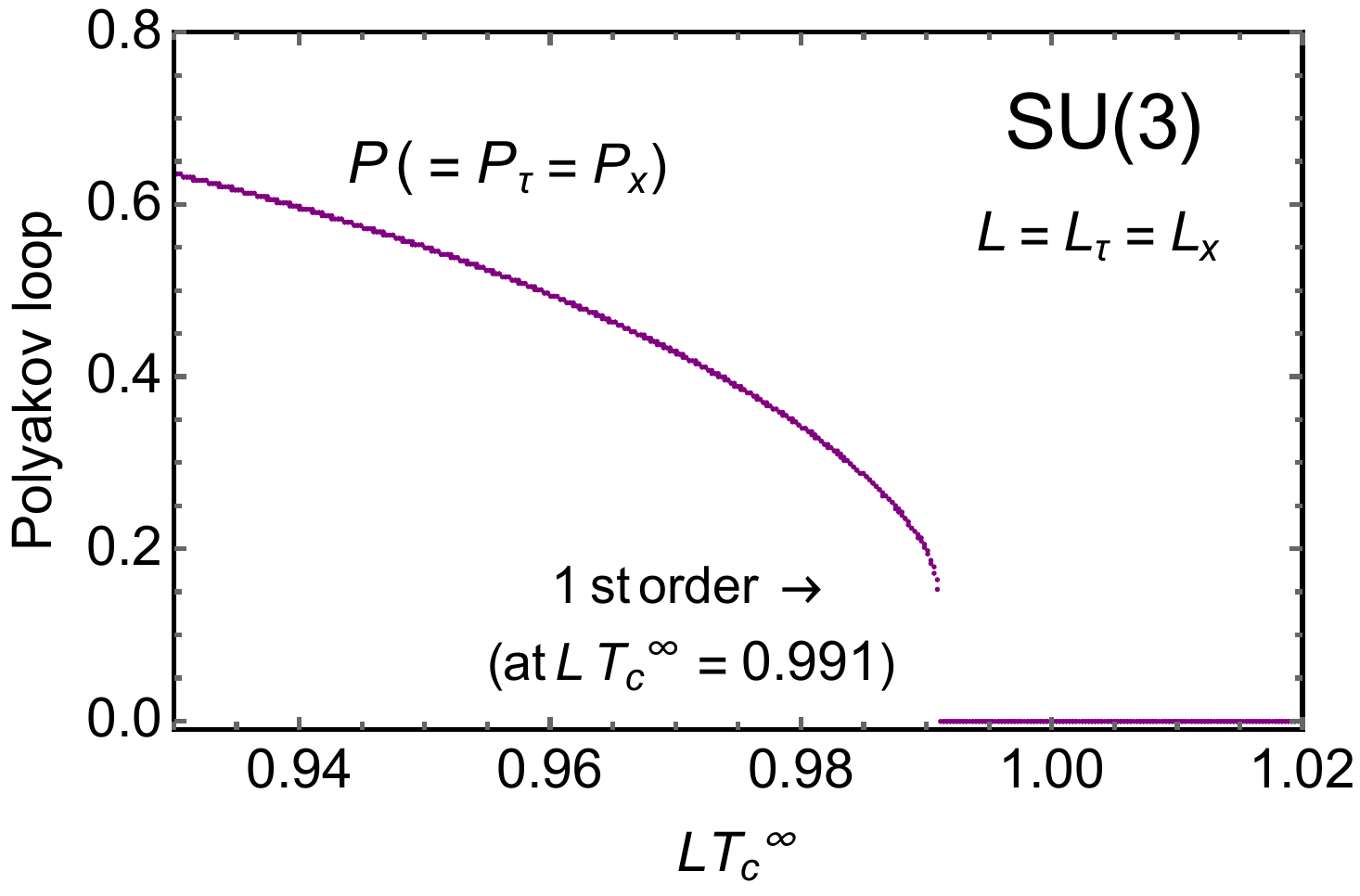}
\caption{$L (\equiv L_\tau=L_x)$ dependence of the Polyakov loop $P(\equiv P_\tau=P_x)$ for $N=3$ in the vicinity of the phase transition point. The first-order transition occurs at $LT^\infty_c\approx 0.991$.}
\label{fig:SymSU3}
\end{figure}

\begin{figure}[hbtp]
\centering
\hspace*{-0.5cm} 
\includegraphics*[scale=0.54]{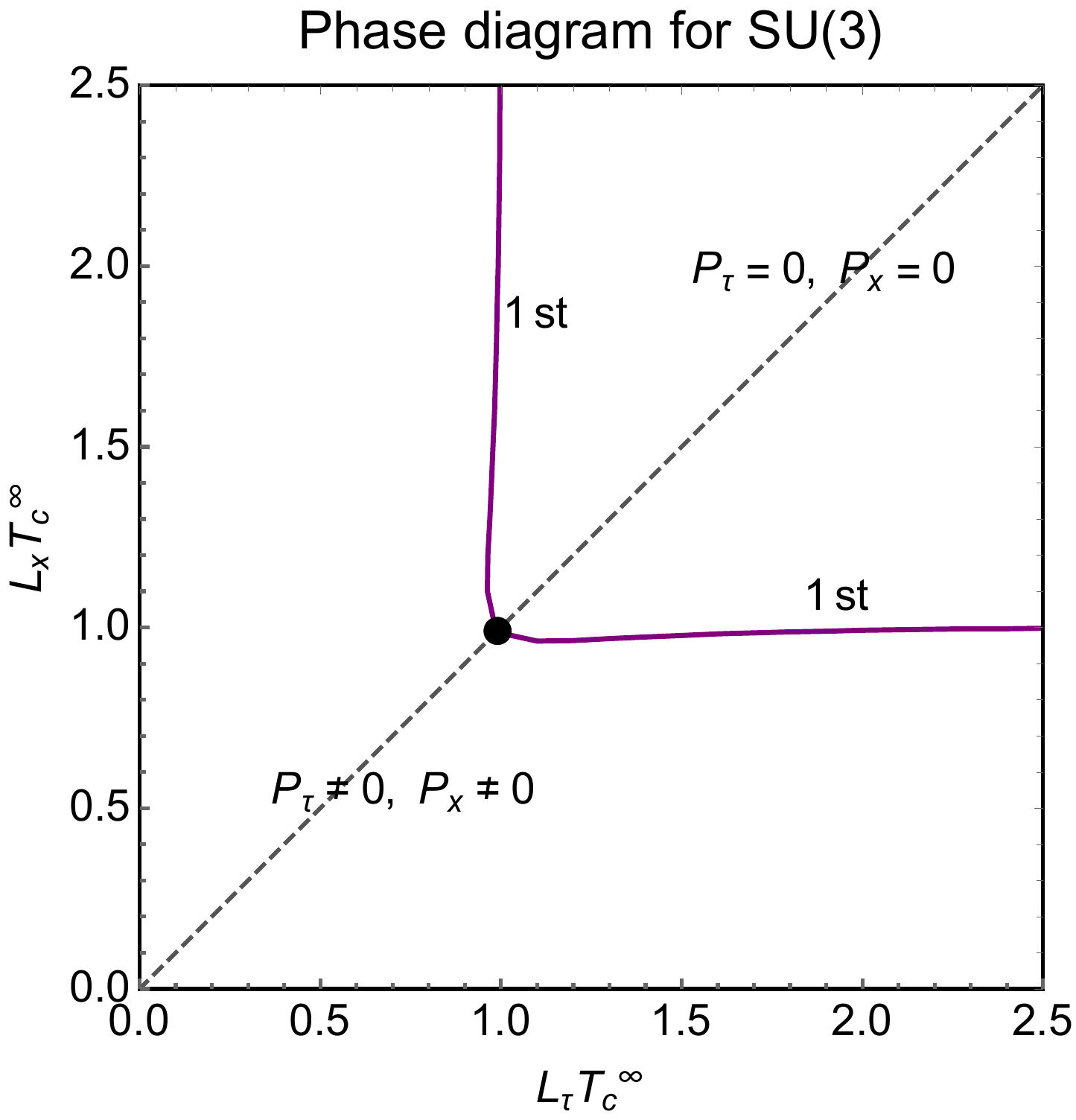}
\caption{Phase diagram on the $L_\tau$--$L_x$ plane for $N=3$. The solid line shows the first-order phase transition.}
\label{fig:PhaseSU3}
\end{figure}

In this subsection we explore the case for $N=2$. In Fig.~\ref{fig:FiniteTSU2} we first show the $L_\tau$ dependence of the Polyakov loops at $L_x\to\infty$ for $N=2$ in order to check the behavior on $\mathbb{S}^1\times \mathbb{R}^3$~\cite{Meisinger:2001cq}. The blue and red curves represent the temporal and spatial Polyakov loops $P_\tau$ and $P_x$, respectively. As can be seen, $P_\tau$ experiences a phase transition. As in Appendix~\ref{sec:GLAnalysis}, it is shown from the Ginzburg-Landau analysis that this is a second-order phase transition with the critical temperature $T_c^\infty=1/((2/3)^{1/3}R) \approx 1/(0.874R)$. The $Z_2^{(\tau)}$ symmetry is broken for $L_\tau T_c^\infty\leq1$ corresponding to the deconfinement. On the other hand, $P_x$ is always zero in $L_x\to\infty$ limit as in Eq.~(\ref{PxLx->inf}).

In Fig.~\ref{fig:3DPsiSU2} we display the $L_\tau$ dependence of $P_\tau$ (top panel) and $P_x$ (bottom panel) for several values of $L_x$. The upper panel indicates that the deconfinement phase transition on $\mathbb{S}^1\times \mathbb{R}^3$ persists even for finite $L_x$. The critical value of $L_\tau$ first decreases with decreasing $L_x$, but it suddenly starts increasing at the symmetric point $L_\tau=L_x$. Meanwhile, the bottom panel shows that a phase with the spontaneous breaking of $Z_2^{(x)}$ symmetry appears at small $L_\tau$, and the critical value of $L_\tau$ for this phase transition becomes larger with decreasing $L_x$. Moreover, $Z_2^{(x)}$ symmetry is mostly broken for sufficiently small $L_x$ ($L_xT_c^\infty\lesssim 0.9$).

While it is analytically shown that the phase transition at $L_x\to\infty$ is of second order, it is difficult to obtain a definite conclusion analytically for finite $L_x$. Our numerical results, however, strongly suggest that the order parameters change continuously and thus the phase transitions are of second order for any value of $L_x$. To see this, in Fig.~\ref{fig:SymSU2} we focus on the symmetric case $L_x=L_\tau=L$ and show the $L$ dependence of the Polyakov loops that behave $P =P_\tau=P_x$ in this case. As shown in Appendix~\ref{sec:GLAnalysis}, provided that the phase transition is of second order it occurs at
\begin{eqnarray}
L T_c^\infty=(2\ln2/\pi)^{1/3}RT_c^\infty=(3\ln2/\pi)^{1/3}\approx0.871\ . \label{AnalSU2}
\end{eqnarray}
The figure suggests that the order parameter becomes nonzero at this point without discontinuity. The analytic solution~(\ref{AnalSU2}) is denoted by the vertical line in the figure.

These results are well summarized as the phase diagram on the $L_\tau$--$L_x$ plane shown in Fig.~\ref{fig:PhaseSU2}. The blue (red) curve separates $P_\tau=0$ and $P_\tau\neq0$ ($P_x=0$ and $P_x\neq0$). The dashed gray line stands for $L_\tau=L_x$, and the black dot represents the transition point on this line.

Finally, let us investigate the fate of the red line that separates $P_x\ne0$ and $P_x=0$ in the limit $(L_x,L_\tau)\to(\infty,0)$. As discussed in Appendix~\ref{sec:GLAnalysis}, provided the second-order transition the limiting behavior of this line is given as in Eq.~(\ref{crit}). In Fig.~\ref{fig:PhaseSU2}, we show Eq.~(\ref{crit}) by the dotted line. The line agrees well for $L T_c^\infty\gtrsim1.3$.

\subsection{$N=3$}
\label{sec:ResultsSU3}

Next, we explore the $N=3$ case. Depicted in Fig.~\ref{fig:FiniteTSU3} is the $L_\tau$ dependence of $P_\tau$ and $P_x$ at $L_x\to\infty$. The blue and red curves represent $P_\tau$ and $P_x$, respectively. One sees from the figure that the deconfined phase is realized for $L_\tau=1/T_c^\infty$, with the critical temperature numerically estimated as $T_c^\infty\approx1/(0.733R)$. Unlike $N=2$, the numerical result shows that $P_\tau$ has a clear discontinuity at this point, which means that the phase transition is of first order. As discussed in Appendix~\ref{sec:GLAnalysis}, this result is analytically confirmed by the Ginzburg-Landau analysis.

In Fig.~\ref{fig:3DPsiSU3} we display the $L_\tau$ dependence of $P_\tau$ (top panel) and $P_x$ (bottom panel) at several $L_x$. At larger $L_x$, the top panel implies that the first-order phase transition of $P_\tau$ observed at $L_x\to\infty$ persists to $L_xT_c^\infty\sim1$. One also finds that both $P_\tau$ and $P_x$ change discontinuously at the identical transition point. The first-order phase transition then ceases to exist at adequately small $L_x$ ($L_xT_c^\infty\lesssim 1$).

To see the order of the phase transition, in Fig.~\ref{fig:SymSU3} we show the $L(=L_\tau=L_x)$ dependence of $P(=P_\tau=P_x)$ for the symmetric case $L_\tau=L_x$ around the transition point. As shown in the figure, $P$ changes discontinuously at $LT_c^\infty\approx0.991$ meaning that the transition is of first order even at the symmetric point. Thus, we conclude that the transition is always of first order for $N=3$.

Finally, we show the phase diagram on the $L_\tau$--$L_x$ plane in Fig.~\ref{fig:PhaseSU3}. As explained above the first-order phase transitions of $P_\tau$ and $P_x$ occur simultaneously. This transition is shown by the purple line. The dashed gray line stands for $L_\tau=L_x$, and the black dot represents the transition point on this line. Note that $P_x$ approaches zero for $L_x\to \infty$ such that the condition in Eq.~(\ref{PxLx->inf}) is satisfied, while the discontinuity of $P_x$ does exist at the first-order transition point.

\begin{figure}[hbtp]
  \centering
  \includegraphics*[width=0.46\textwidth]{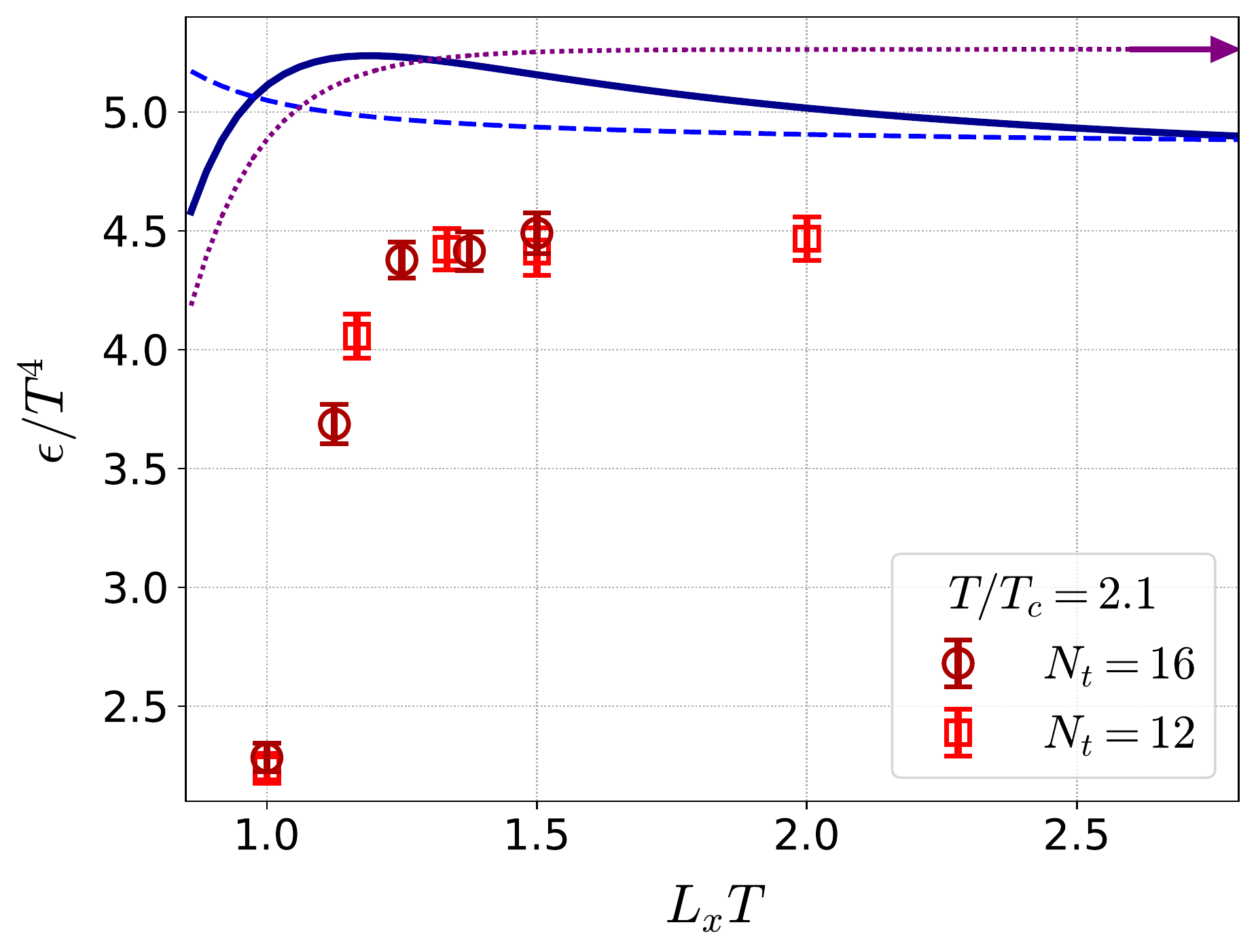}
  \includegraphics*[width=0.46\textwidth]{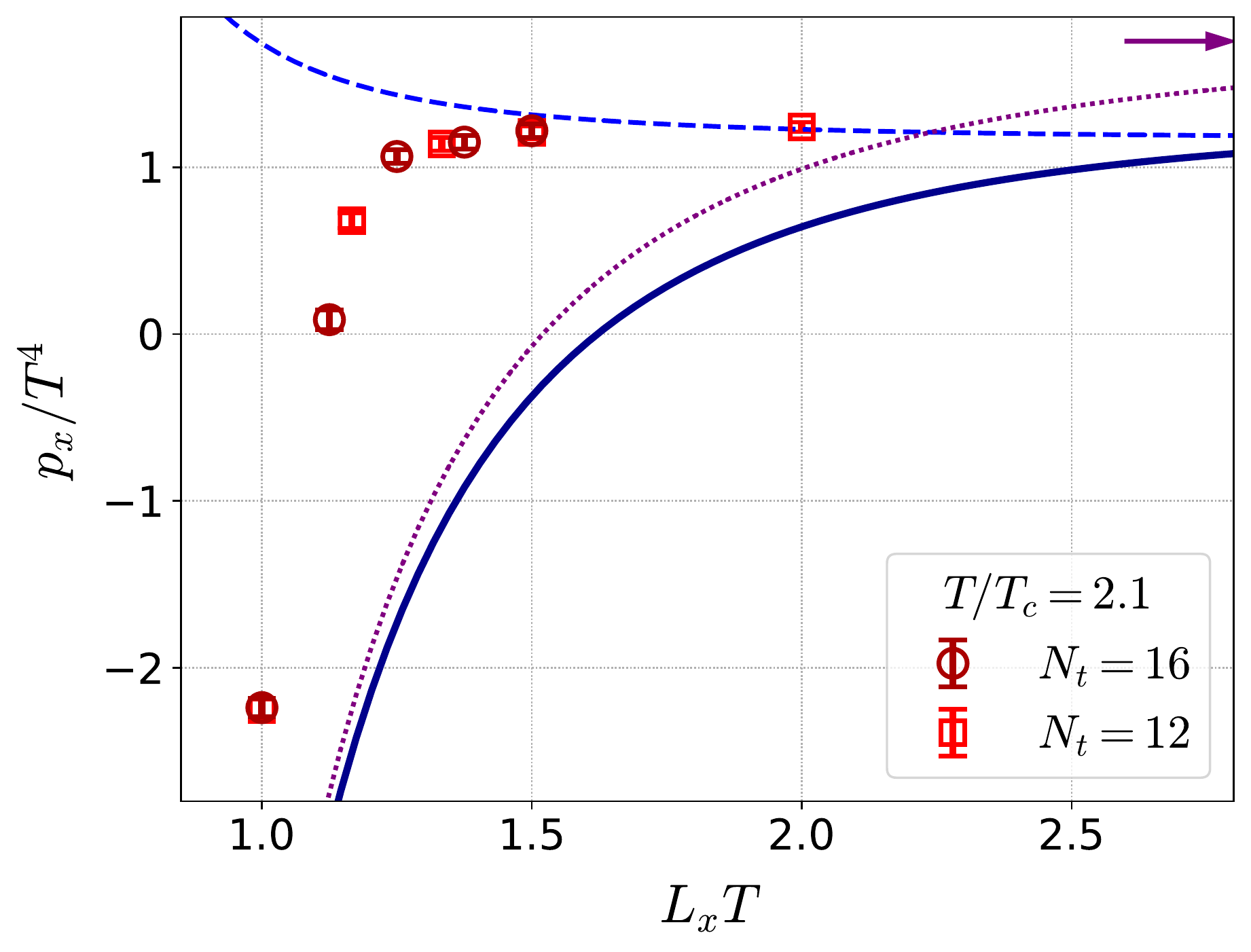}
  \includegraphics*[width=0.46\textwidth]{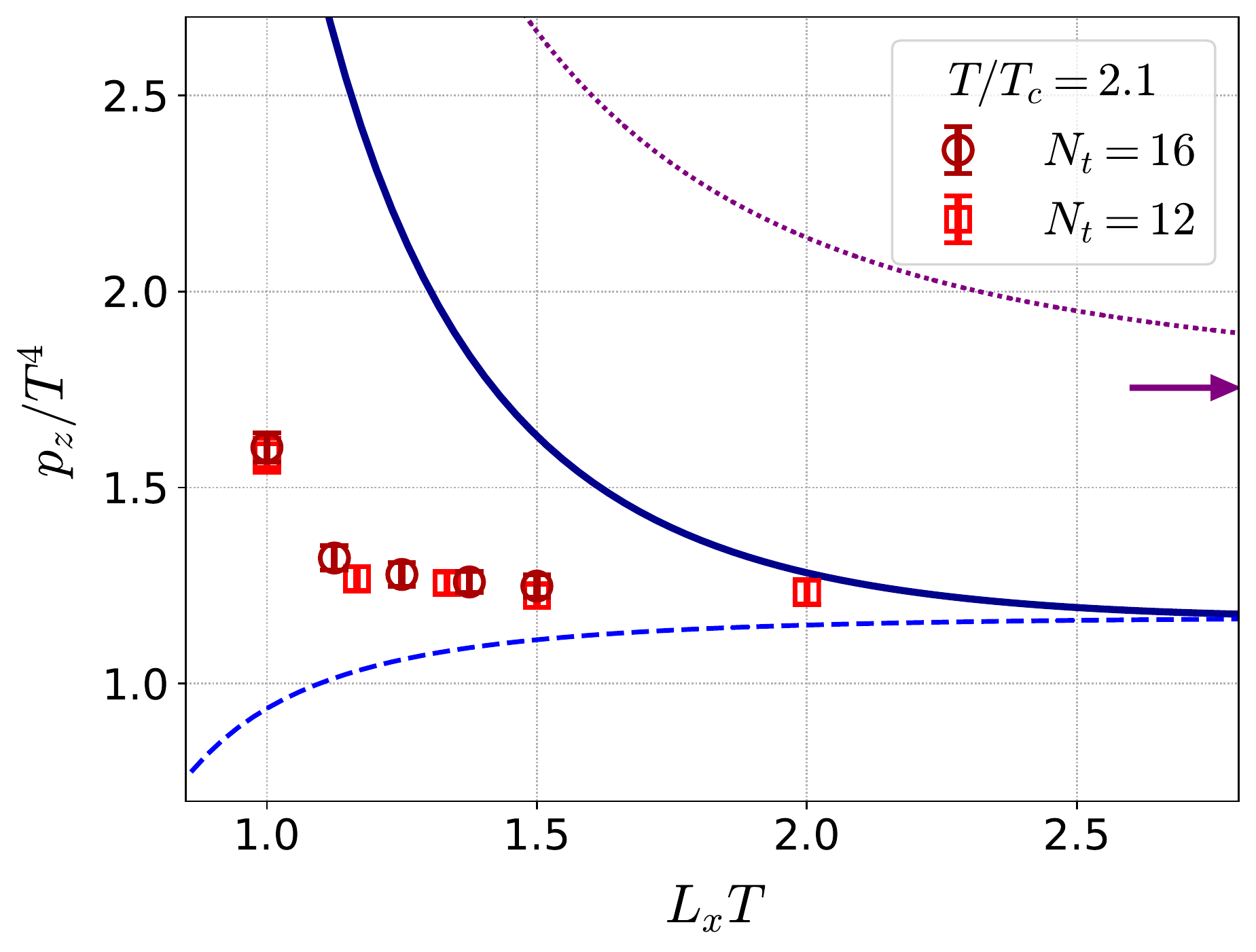}
  \caption{Comparison of energy density $\epsilon$ and pressures $p_x$ and $p_z$ between our model (solid lines) and the lattice data in Ref.~\cite{Kitazawa:2019otp} with $N_t=16$ (circles) and $12$ (squares). The dashed lines show the results with fixed values of $P_x$ and $P_t$ given in the $L_x\to\infty$ limit. The dotted lines are the results in the massless free theory and their limiting values for $L_xT\to\infty$ are shown by the arrows.}
\label{fig:Comparison}
\end{figure}

\begin{figure}[hbtp]
  \centering
  \includegraphics*[width=0.46\textwidth]{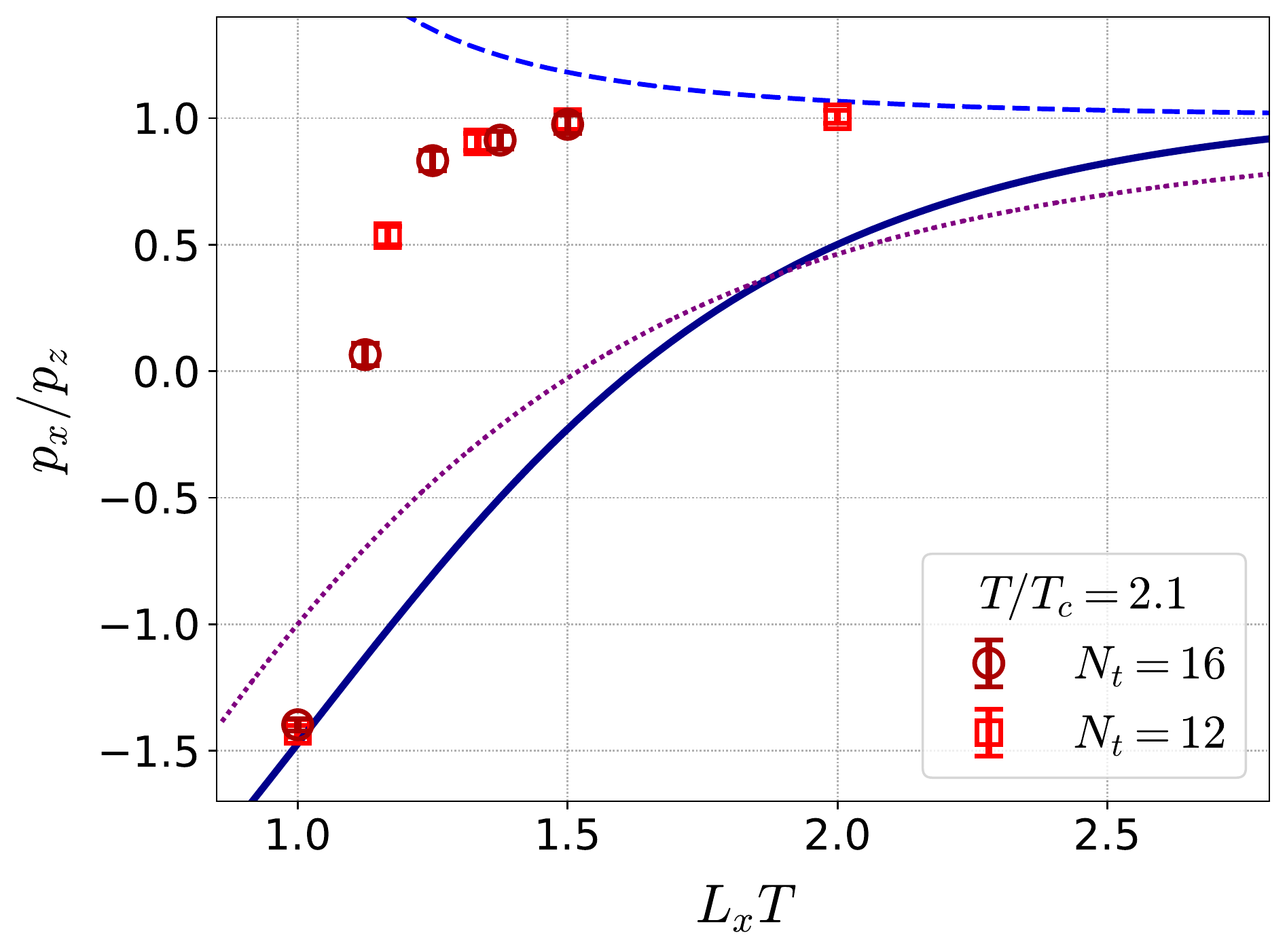}
  \caption{Pressure ratio $p_x/p_z$. The meanings of the symbols are the same as Fig.~\ref{fig:Comparison}.} 
\label{fig:pxpz}
\end{figure}

\section{Thermodynamics}
\label{sec:Thermodynamics}

In this section, we investigate $L_\tau$ and $L_x$ dependence of thermodynamic quantities on $\mathbb{T}^2\times\mathbb{R}^2$ and study the impact of $P_\tau$ and $P_x$ on them.

An important feature of the thermodynamics on $\mathbb{T}^2\times\mathbb{R}^2$ is that the pressure becomes anisotropic because of the violation of rotational symmetiry due to the BC~\cite{Kitazawa:2019otp}. In such systems, the stress tensor $\sigma_{ij}$ with $i,j=x,y,z$ should be used in place of the pressure to represent the force acting on the surface. The stress tensor is equivalent with the spatial components of the energy-momentum tensor $T^\mu_\nu$ up to the overall sign. The spatial components of $T^\mu_\nu$, and thus $\sigma_{ij}$, have off-diagonal components in general. However, in our setting that imposes a PBC along $x$ direction, due to the parity symmetry the energy-momentum tensor in the Minkowski space is given by the diagonal form as 
\begin{eqnarray}
  T^\mu_\nu = {\rm diag} (\epsilon,p_x,p_y,p_z) \ ,
  \label{Tmunu}
\end{eqnarray}
where $\epsilon$ is the energy density and $p_i$ represents the pressure for each direction. Because of the rotational symmetry around $x$ axis $p_y=p_z$ is satisfied, but $p_x$ can be different from $p_y$ and $p_z$.

The values of $\epsilon$, $p_x$ and $p_z$ in our model are obtained from the free energy as 
\begin{eqnarray}
\epsilon &=& \frac{L_\tau}{ {\cal V}}\frac{\partial}{\partial L_\tau} {\cal V}f \ , \nonumber\\
p_x &=& -\frac{L_x}{ {\cal V}}\frac{\partial}{\partial L_x} {\cal V}f \ , \nonumber\\
p_z &=& -\frac{L_z}{ {\cal V}}\frac{\partial}{\partial L_z} {\cal V}f \ , 
\end{eqnarray}
where we have introduced the Euclidean four dimensional volume ${\cal V}=L_\tau L_xL_yL_z$ and temporary assumed that the lengths for $y$ and $z$ directions, $L_y$ and $L_z$, are finite.

In Fig.~\ref{fig:Comparison}, we show the $L_xT(=L_x/L_\tau)$ dependence of $\epsilon$, $p_x$ and $p_z$ obtained in our model for $N=3$ at $T/T_c^\infty=1/(L_\tau T_c^\infty)=2.10$ by the solid blue lines. In the figure, same quantities calculated in the massless free theory are shown by the dotted lines for comparison. Their limiting values for $L_xT\to\infty$ are shown by the arrows. In the figure, we also plot these quantities in $SU(3)$ YM theory on $\mathbb{T}^2\times\mathbb{R}^2$ for the same $T$ obtained in Ref.~\cite{Kitazawa:2019otp} for comparison. The circle and square symbols show the results obatined for $N_\tau=16$ and $12$, respectively, where $N_\tau$ is the number of lattice sites along the temporal direction which is related to the lattice spacing $a$ as $N_\tau=(aT)^{-1}$.

From Fig.~\ref{fig:Comparison}, one sees that our model results do not show good agreement with the lattice ones even qualitatively. In particular, while the lattice data show that the values of $p_x$ and $p_z$ hardly change from the one in the $L_x\to\infty$ limit for $L_xT\gtrsim1.5$, this behavior is not reproduced in the model calculation. Similar results are obtained for other values of $T/T_c$ investigated in Ref.~\cite{Kitazawa:2019otp}. Figure~\ref{fig:pxpz} compares these results in term of the pressure ratio $p_x/p_z$. From the figure one sees that the ratio obtained in our model is closer to unity than the massless-free result for $L_xT\gtrsim2.0$, but the modification is not enough to reproduce the lattice results.

These results, however, show that the non-trivial expectation values of $P_\tau$ and $P_x$ give non-negligible contribution to the behavior of thermodynamics on $\mathbb{T}^2\times\mathbb{R}^2$. In order to gain insights into their effects, we perform an additional analysis, where the values of $P_x$ and $P_\tau$ are fixed by hand to those at $L_x\to\infty$; $P_\tau\approx0.973$ and $P_x=0$. The obtained results for $\epsilon$, $p_x$ and $p_z$ are shown in Figs.~\ref{fig:Comparison} and~\ref{fig:pxpz} by the dashed lines. As can be seen, these results are insensitive to $L_xT$ and give consistent behavior with the lattice data for $L_xT\gtrsim1.5$. These results show that $P_x$ and $P_\tau$ affect thermodynamic quantities on $\mathbb{T}^2\times\mathbb{R}^2$ significantly. Therefore, although we have failed in reproducing the lattice data in the present study with a simple model, the modification of the model, especially the potential term, would give a consistent result with the lattice data. Such a description of the lattice results on $\mathbb{T}^2\times\mathbb{R}^2$ will in turn give us deeper understanding on the non-perturbative aspects of YM theory near $T_c^\infty$ not only on $\mathbb{T}^2\times\mathbb{R}^2$ but also $\mathbb{S}^1\times\mathbb{R}^3$.

\section{Conclusion and outlook}
\label{sec:Conclusions}

In this paper, we have investigated the phase structure and thermodynamics of the pure Yang-Mills (YM) theory on $\mathbb{T}^2\times\mathbb{R}^2$ with the PBC by means of an effective model. The model has two Polyakov loops along the compactified directions, $P_\tau$ and $P_x$, as the order parameters and thus is capable of describing the phase transitions associated with two $Z_N$ symmetries. As a first investigation of such an effective model, we have employed a simple form for the potential term $f_{\rm pot}$ given by an extension of Ref.~\cite{Meisinger:2001cq}. We have found that a rich phase structure on the $L_\tau$--$L_x$ plane can manifest itself due to two phase transitions, and the phase structure is qualitatively dependent on $N$.

The energy density and anisotropic pressure are also calculated in the model and are compared with the lattice results in Ref.~\cite{Kitazawa:2019otp}. Although we have found that our model fails in reproducing the lattice results, we have also found that thermodynamics on $\mathbb{T}^2\times\mathbb{R}^2$ is sensitive to $P_\tau$ and $P_x$ in the model. Therefore, while the present model with a simple ansatz is not satisfactory, the modification of the model would be able to reproduce the lattice results. The analysis in the previous section that fixes the values of $P_\tau$ and $P_x$ by hand would be used for a guide for such a study. We also note that the model can be improved toward other directions, for example, introduction of the quasi-particle mass of the gauge field and other mean fields~\cite{Gorenstein:1995vm,Peshier:1995ty,Carter:1998ti,Sannino:2002wb,Brau:2009mp,Begun:2010eh,Ruggieri:2012ny,Sasaki:2012bi,Sasaki:2013xfa} especially mimicking the magnetic condensates in the deconfined phase~\cite{Agasian:2003yw}. Comparison with the stress tensor obtained in the AdS/CFT correspondence~\cite{Balasubramanian:1999re,Myers:1999psa} is also interesting to investigate the role of the strong-coupling nature. 

The investigation of the field theory with the BC can also be extended to various directions. An example is the use of other BCs such as the anti-periodic BC in place of the PBC. Although we have limited our attention only to $N=2,3$, the study of $N$ dependence for $N\ge4$ is a straightforward extension of the present study. Similar analysis in QCD with dynamical fermions is another important subject. Since the measurement of thermodynamics in QCD on $\mathbb{T}^2\times\mathbb{R}^2$ is possible using the technique developed in Refs.~\cite{Makino:2014taa,Taniguchi:2016ofw,Taniguchi:2020mgg}, the comparison with the lattice data is possible.

\section*{Acknowledgement}

The authors thank Kouji Kashiwa, Makoto Natsuume and Yuya Tanizaki for giving us useful comments. This work is supported by the RIKEN special postdoctoral researcher program (D.~S.), and JSPS KAKENHI Grant Numbers JP19H05598, JP20H01903, JP22K03619 (M.~K.).

\appendix

\section{DERIVATION OF EQS.~(\ref{FPertSU2}) and (\ref{FPertSU3})}
\label{sec:FPert}

In this appendix we show derivation of Eqs.~(\ref{FPertSU2}) and (\ref{FPertSU3}).

For $N=2$, choosing the phase variables as Eq.~(\ref{AnsatzSU2}), the non vanishing $(\Delta\theta_c)_{jk}$ reads
\begin{eqnarray}
(\Delta\theta_c)_{12}=-(\Delta\theta_c)_{21}=2\phi_c\  . \label{DeltaSU2}
\end{eqnarray}
Hence, the one-loop free energy~(\ref{FPert}) is of the form
\begin{eqnarray}
f_{\rm pert}&=& \frac{1}{L_\tau L_x}\sum_{l_\tau,l_x} \int\frac{d^2p_L}{(2\pi)^2} {\rm ln}(\omega_\tau^2+\omega_x^2+{\bm p}_L^2) \nonumber\\
&+& \frac{2}{L_\tau L_x}\sum_{l_\tau,l_x} \int\frac{d^2p_L}{(2\pi)^2} \nonumber\\
&\times& {\rm ln}\left[\left(\omega_\tau-\frac{2\phi_\tau}{L_\tau}\right)^2+\left(\omega_x-\frac{2\phi_x}{L_x}\right)^2+{\bm p}_L^2\right]  \ , \nonumber\\ \label{FPertSU2P}
\end{eqnarray}
with ${\bm p}_L=(p_y,p_z)$, $\omega_\tau=2\pi l_\tau$ and $\omega_x=2\pi l_x$ ($l_\tau$ and $l_x$ are integers). Here, with the help of regularization procedures provided in detail in Appendix~\ref{sec:ZetaFunction}, the one-loop free energy $f_{\rm pert}$ is appropriately regularized to be
\begin{eqnarray}
f_{\rm pert} &=& -\frac{1}{2\pi^2}\overline{\sum_{l_\tau, l_x}}\frac{m^2}{X_{l_\tau,l_x}^2} K_{2}\left(m X_{l_\tau,l_x}\right) \nonumber\\
&\times& \left(1+ 2{\rm e}^{2i\phi_\tau l_\tau}{\rm e}^{2i\phi_xl_x} \right)\Big|_{m\to0}\ .  \label{FPertReg}
\end{eqnarray}
The symbol $\overline{\sum}_{l_\tau, l_x}$ in Eq.~(\ref{FPertReg}) is defined such that the summation does not include contributions from $(l_\tau,l_x)=(0,0)$. $K_y(x)$ in Eq.~(\ref{FPertReg}) is the modified Bessel function of the second kind, and we have defined
\begin{eqnarray}
X_{l_\tau,l_x} \equiv \sqrt{(l_\tau L_\tau )^2+(l_xL_x)^2} \ .
\end{eqnarray}
It should be noted that we have included the mass of the gauge field $m$ for convenience, but soon we take $m\to0$. That is, using $K_2(x) \approx 2/x^2$ for $x\approx0$, the free energy~(\ref{FPertReg}) is reduced to
\begin{eqnarray}
f_{\rm pert} &=& -\frac{1}{\pi^2}\overline{\sum_{l_\tau, l_x}}\frac{1+ 2{\rm e}^{2i\phi_\tau l_\tau}{\rm e}^{2i\phi_xl_x}}{X_{l_\tau,l_x}^4} \nonumber\\
&=& -\frac{2}{\pi^2}\Bigg[\sum_{l_\tau=1}^\infty\frac{1+2 \cos(2\phi_\tau l_\tau) }{X_{l_\tau,0}^4} \nonumber\\
&+& \sum_{l_x=1}^\infty\frac{1+2\cos(2\phi_x l_x) }{X_{0,l_x}^4}  \nonumber\\
&+& 2\sum_{l_\tau,l_x=1}^\infty\frac{1+2 \cos(2\phi_\tau l_\tau)\cos(2\phi_x l_x)}{X_{l_\tau,l_x}^4}  \Bigg] \ . \label{FPertM0}
\end{eqnarray}
The first and second lines in the second equality in Eq.~(\ref{FPertM0}) are separate contributions from the compactified $\tau$ and $x$ directions, respectively, while the third one contains interplays between them leading to a rich phase structure of the $Z_2^{(\tau)}\times Z_2^{(x)}$ on $\mathbb{T}^2\times\mathbb{R}^2$. The separate contributions are reduced to rather familiar forms. In fact, the summations are rewritten by the Bernoulli polynomial by the following infinite series:
\begin{eqnarray}
B_{2n}(x) &=& (-1)^{n+1}\frac{2(2n)!}{(2\pi)^{2n}}\sum_{k=1}^\infty\frac{\cos(2\pi k x)}{k^{2n}} \ ,
\end{eqnarray}
with $0\leq x\leq 1$, leading to
\begin{eqnarray}
f_{\rm pert} &=& \frac{2\pi^2}{3L_\tau^4}\Big[B_4(0)+2B_4(\phi_\tau/\pi)\Big]  \nonumber\\
&+& \frac{2\pi^2}{3L_x^4}\Big[B_4(0)+2B_4(\phi_x/\pi)\Big] \nonumber\\
&-& \frac{4}{\pi^{2}}\sum_{l_\tau,l_x=1}^\infty\frac{1+2\cos(2\phi_\tau l_\tau)\cos(2\phi_xl_x) }{X_{l_\tau,l_x}^4} \ . \label{FPertBessel}
\end{eqnarray}
Therefore, remembering the elementary expression of the Bernoulli polynomial
\begin{eqnarray}
B_4(x) = x^2(x-1)^2-\frac{1}{30}\ ,
\end{eqnarray}
we finally arrive at Eq.~(\ref{FPertSU2}).

For $N=3$, the parametrization in Eq.~(\ref{AnsatzSU3}) enables us to rewrite the one-loop free energy~(\ref{FPert}) into
\begin{eqnarray}
f_{\rm pert} &=& \frac{2}{L_\tau L_x}\sum_{l_\tau,l_x} \int\frac{d^2p_L}{(2\pi)^2} {\rm ln}(\omega_\tau^2+\omega_x^2+{\bm p}_L^2) \nonumber\\
&+& \frac{4}{L_\tau L_x}\sum_{l_\tau,l_x} \int\frac{d^2p_L}{(2\pi)^2}  \nonumber\\
&\times& {\rm ln}\left[\left(\omega_\tau-\frac{\phi_\tau}{L_\tau}\right)^2+\left(\omega_x-\frac{\phi_x}{L_x}\right)^2+{\bm p}_L^2\right] \nonumber\\
&+&\frac{2}{L_\tau L_x}\sum_{l_\tau,l_x}  \int\frac{d^2p_L}{(2\pi)^2} \nonumber\\
&\times& {\rm ln}\left[\left(\omega_\tau-\frac{2\phi_\tau}{L_\tau}\right)^2+\left(\omega_x-\frac{2\phi_x}{L_x}\right)^2+{\bm p}_L^2\right] \ .\nonumber\\
\end{eqnarray}
Thus, tracing a similar evaluation from Eq.~(\ref{FPertReg}) to Eq.~(\ref{FPertBessel}), one can straightforwardly obtain Eq.~(\ref{FPertSU3}).

Next, we take a closer look at the $L_x/L_\tau\to\infty$ limit of $f_{\rm pert}$ focusing on the case $N=2$; the generalization to arbitrary $N$ is straightforward. As discussed in the text, the leading term in this limit is proportional to $1/L_\tau^4$ and it reproduces the result in Ref.~\cite{Meisinger:2001cq}. The subleading term with respect to $r=L_\tau/L_x$ comes from the double-sum term in Eq.~(\ref{FPertSU2}) or (\ref{FPertBessel}). The expansion of this term with respect to $r$ is cumbersome because a simple Taylor expansion of the denominator leads to divergent series. To suppress the divergence, one may first rewrite the double sum as 
\begin{eqnarray}
  &&\sum_{l_\tau,l_x=1}^\infty \frac{1+2\cos(2\phi_\tau l_\tau)\cos(2\phi_xl_x) }{ X_{l_\tau,l_x}^4}
  \nonumber \\
  &&= 
  \frac1{L_\tau^4} \sum_{l_\tau,l_x=1}^\infty
  \frac{1+2\cos(2\phi_\tau l_\tau)\cos(2\phi_xl_x) }{ (l_\tau^2 + l_x^2/r^2)^2}\ .
  \label{sum1}
\end{eqnarray}
Here, we show a useful identity  
\begin{eqnarray}
  &&\sum_{l=1}^\infty \frac{ \cos(2\phi l) }{ ( l^2+C^2 )^2 }
  = \frac12 \sum_{l=-\infty}^\infty \frac{e^{2i\phi l}}{(l^2+C^2)^2} - \frac1{2C^4}
  \nonumber \\
  &&= \pi \sum_{m=-\infty}^\infty \frac{1+2C|\pi m+\phi|}{4C^3} e^{-2C|\pi m+\phi|} - \frac1{2C^4} \ , \nonumber\\
  \label{sum2}
\end{eqnarray}
which can be derived by the Poisson summation formula $\sum_{n=-\infty}^\infty F(n) = \sum_{m=-\infty}^\infty \tilde{F}(m)$ for $\tilde{F}(k)=\int_{-\infty}^\infty dx e^{i 2\pi kx} F(x)$ with $F(x)=e^{2i\phi x}(x^2 + C^2)^{-2}$. For positive and sufficiently large $C$, the first term in Eq.~(\ref{sum2}) is negligible due to the exponential term unless $\phi=0$ so that the term is of the order $C^{-4}$ for $\phi\ne0$. Only when $\phi=0$, non-vanishing contribution survives from $m=0$ and Eq.~(\ref{sum2}) becomes of order $C^{-3}$. Applying the identity~(\ref{sum2}) to the summation with respect to $l_\tau$ in Eq.~(\ref{sum1}) and focusing on a large value of $C=l_x/r$ together with the above order estimates, one can show that Eq.~(\ref{sum1}) starts from ${\cal O}(r^3)$ where the remaining summation for $l_x$ is converging. Therefore, the counting in Eq.~(\ref{Finf_exp}) is derived.

\section{REGULARIZATION BY THE EPSTEIN-HURWITZ ZETA FUNCTION}
\label{sec:ZetaFunction}

When evaluating the one-loop free energy, we encounter the following function:
\begin{eqnarray}
{\cal I} &\equiv& \frac{1}{L_\tau L_x}\sum_{l_\tau,l_x}\int\frac{d^2p_L}{(2\pi)^2} \nonumber\\
&\times& \ln\Big[(\omega_\tau-a_\tau)^2+(\omega_x-a_x)^2+{\bm p}_L^2+ m^2\Big]\ . \label{IDef}
\end{eqnarray}
The function~(\ref{IDef}) includes UV divergences. In this appendix, we show our method to regularize the divergences by means of the inhomogeneous and generalized Epstein-Hurwitz zeta function together with the dimensional regularization~\cite{Elizalde:1995hck}.

To begin with, we perform the momentum integrals in Eq.~(\ref{IDef}) as
\begin{eqnarray}
{\cal I} &=& - \frac{1}{L_\tau L_x}\sum_{l_\tau,l_x}\int\frac{d^dp_L}{(2\pi)^d}  \nonumber\\
&\times&\frac{\partial}{\partial s} \left\{\frac{1}{[(\omega_\tau-a_\tau)^2+(\omega_x-a_x)^2+{\bm p}_L^2 + m^2]^s}\right\}\Big|_{s\to0} \nonumber\\
&=& - \frac{1}{L_\tau L_x}\frac{1}{(4\pi)^{d/2}} \frac{\partial}{\partial s}\left\{ \frac{\Gamma[s-\frac{d}{2}]E(s)}{\Gamma[s]}\right\}\Bigg|_{s\to0} \ , \label{IDerivative}
\end{eqnarray}
where $d=2-\epsilon$ and $E(s)$ is the inhomogeneous and generalized Epstein-Hurwitz zeta function of the form
\begin{eqnarray}
E(s) &=& \sum_{l_\tau,l_x} \left[\frac{1}{(\omega_\tau-a_\tau)^2+(\omega_x-a_x)^2+m^2} \right]^{s-\frac{d}{2}} \ . \nonumber\\
\end{eqnarray}
Although the summations in $E(s)$ are diverging when we take $s\to0$, we utilize the following identity to rearrange the summations in terms of the modified Bessel function of the second kind:
\begin{eqnarray}
&&\sum_{n_1,\cdots, n_N}\left[\frac{1}{w_1(n_1-\alpha_1)^2+\cdots +w_N(n_N-\alpha_N)^2+c^2}\right]^\nu \nonumber\\
&=& \frac{\pi^{N/2}}{\sqrt{w_1\cdots w_N}}\frac{\Gamma\left[\nu-\frac{N}{2}\right]}{\Gamma[\nu]}|c|^{N-2\nu} +\frac{\pi^\nu}{\sqrt{w_1\cdots w_N}}\frac{2}{\Gamma[\nu]} \nonumber\\
&\times&\overline{\sum_{n_1,\cdots,n_N}}  {\rm exp}\big[2\pi i(n_1\alpha_1+\cdots +n_N\alpha_N)\big] \nonumber\\
&\times&  |c|^{\frac{N}{2}-\nu}\left[\frac{n_1^2}{w_1}+\cdots+\frac{n_N^2}{w_N}\right]^{\frac{1}{2}\left(\nu-\frac{N}{2}\right)} \nonumber\\
&\times& K_{\frac{N}{2}-\nu}\left(2\pi |c|\left[\frac{l_1^2}{w_1}+\cdots+\frac{l_N^2}{w_N}\right]^{1/2}\right) \ , \label{EHFormula}
\end{eqnarray}
with $n_1,\cdots, n_N$ being integers. The symbol $\overline{\sum}_{n_1,\cdots, n_x}$ in Eq.~(\ref{EHFormula}) is defined such that the summation does not include contributions from $(n_1,\cdots,n_N)=(0,\cdots,0)$. The rearrangement is useful to separate the diverging parts. Thanks to Eq.~(\ref{EHFormula}), the $\Gamma[s-\frac{d}{2}]E(s)$ part in Eq.~(\ref{IDerivative}) becomes finite as long as we keep $\epsilon$ finite even when $s\to0$ is taken. Here, using $\Gamma[s] \approx 1/s-\gamma_E+{\cal O}(s)$ with the Euler's constant $\gamma_E=0.577\cdots$, we can get 
\begin{eqnarray}
\frac{\partial}{\partial s}\left\{ \frac{\Gamma[s-\frac{d}{2}]E(s)}{\Gamma[s]}\right\} &=&\frac{\partial}{\partial s}\left\{s\Gamma\left[s-\frac{d}{2}\right]E(s)+\cdots\right\}\nonumber\\
&\overset{s\to0}{=}&\Gamma\left[-\frac{d}{2}\right] E(0)\ .  \label{NIdentity}
\end{eqnarray}
Therefore, from Eqs.~(\ref{EHFormula})~and~(\ref{NIdentity}), the function~(\ref{IDerivative}) is regularized and evaluated as
\begin{eqnarray}
{\cal I} &=& - \frac{1}{L_\tau L_x}\frac{1}{(4\pi)^{d/2}}\Gamma\left[-\frac{d}{2}\right]E(0) \nonumber\\
 &=& \frac{1}{(4\pi)^{(d+2)/2}}\Gamma\left[-\frac{d+2}{2}\right]\left(\frac{1}{m^2}\right)^{-(d+2)/2} \nonumber\\
 &+& \frac{1}{2\pi^2}\overline{\sum_{l_\tau,l_x}}\frac{m^2}{X_{l_\tau,l_x}^2}{\rm e}^{i(l_\tau L_\tau a_\tau+l_xL_xa_x)}K_{2}(mX_{l_\tau,l_x})\ .\nonumber\\ \label{IRegulate}
\end{eqnarray}
Remembering $d=2-\epsilon$, the first term in Eq.~(\ref{IRegulate}) coincides with the result in vacuum within the ordinary dimensional regularization, implying that the UV divergences are appropriately regularized. Thus, the second term in Eq.~(\ref{IRegulate}) is well-defined and must be understood as contributions from PBC. Obviously, such a clear separation is successfully done by use of the identity~(\ref{EHFormula}).


\section{GINZBURG-LANDAU ANALYSIS}
\label{sec:GLAnalysis}

Here, we make use of the Ginzburg-Landau theory to delineate the order of the phase transitions found in Sec.~\ref{sec:PhaseDiagram}.

First, we focus on the free energy for $N=2$ where the free energy per unit volume is given by Eqs.~(\ref{FPertSU2}) and~(\ref{FHaarSU2}).
We change the phase variables $\phi_\tau$ and $\phi_x$ to
\begin{eqnarray}
\psi_\tau \equiv \frac{\pi}{2}-\phi_\tau \ , \ \ \psi_x \equiv \frac{\pi}{2}-\phi_x\ , \label{PsiDef}
\end{eqnarray}
such that $\psi_\tau=0$ and $\psi_x=0$ correspond to the $Z_{2}^{(\tau)}$ and $Z_{2}^{(x)}$ symmetric phases, respectively. Substituting Eq.~(\ref{PsiDef}) to Eqs.~(\ref{FPertSU2}) and~(\ref{FHaarSU2}), $f$ is expressed in terms of $\psi_\tau$ and $\psi_x$ as
\begin{eqnarray}
f &=& -\frac{\pi^2}{15L_\tau^4} + \frac{1}{12\pi^2L_\tau^4}(\pi^2-4\psi_\tau^2)^2 \nonumber\\
&-& \frac{\pi^2}{15L_x^4} + \frac{1}{12\pi^2L_x^4}(\pi^2-4\psi_x^2)^2 \nonumber\\
&-&\frac{4}{\pi^{2}}\sum_{l_\tau,l_x=1}^\infty\frac{1+2(-1)^{l_\tau+l_x}\cos(2\psi_\tau l_\tau)\cos(2\psi_xl_x)}{X_{l_\tau,l_x}^4}  \nonumber\\
&-& \frac{1}{L_\tau R^3}{\rm ln}({\rm cos}^2\psi_\tau) -  \frac{1}{L_xR^3}{\rm ln}({\rm cos}^2\psi_x)\ . \label{FPsiSU2}
\end{eqnarray}
By expanding the free energy~(\ref{FPsiSU2}) around $\psi_c=0$, one can investigate the critical behavior.

When taking $L_x\to\infty$, the second and third lines of Eq.~(\ref{FPsiSU2}) are negligible. The free energy thus includes only $\psi_\tau$ and can be expanded as
\begin{eqnarray}
f &=& \frac{\pi^2}{60L_\tau^4}+\left(\frac{1}{L_\tau R^3}-\frac{2}{3L_\tau^4}\right)\psi_\tau^2  \nonumber\\
&+& \left(\frac{4}{3\pi^2L_\tau^4}+\frac{1}{6L_\tau R^3}\right)\psi_\tau^4 + \frac{2}{45L_\tau R^3}\psi_\tau^6 + \cdots \ .\nonumber\\  \label{FPsiSU2Ex}
\end{eqnarray}
In this expansion, contributions of ${\cal O}(\psi_\tau^6)$ stems from the logarithms in the last line in Eq.~(\ref{FPsiSU2}), and thanks to cosines there remain only terms having even powers of $\psi_\tau$ with positive coefficients. Hence, only the coefficient of $\psi_\tau^2$ in Eq.~(\ref{FPsiSU2Ex}) can change its sign. For this reason we can conclude that the second-order phase transition takes place at 
\begin{eqnarray}
L_\tau = \left(\frac{2}{3}\right)^{1/3}R \approx 0.874 R\ .
\end{eqnarray}

Next we consider the symmetric case $L\equiv L_\tau=L_x$. Assuming that $\psi\equiv \psi_\tau=\psi_x$, the free energy~(\ref{FPsiSU2}) is expanded as
\begin{eqnarray}
f &=&\left( \frac{\pi^2}{30L^4}-\frac{4}{\pi^2L^4}\sum_{l_\tau,l_x}\frac{1+2(-1)^{l_\tau+l_x}}{(l_\tau^2+l_x^2)^2}\right) \nonumber\\
&+&\left(\frac{2}{LR^3}-\frac{4}{3L^4}+\frac{16}{\pi^2L^4}\sum_{l_\tau,l_x}\frac{(-1)^{l_\tau+l_x}}{l_\tau^2+l_x^2}\right)\psi^2 \nonumber\\
&+& \cdots\ . \label{FPsiSU2ExL}
\end{eqnarray}
Using
\begin{eqnarray}
  \sum_{l_\tau,l_x}\frac{(-1)^{l_\tau+l_x}}{l_\tau^2+l_x^2}
  = \frac{ \pi^2 - 3 \pi \ln 2}{12} \ ,
\end{eqnarray}
the coefficient of $\psi^2$ in Eq.~(\ref{FPsiSU2ExL}) is calculated to be
\begin{eqnarray}
  \frac2{L^4}\Big( -\frac2\pi \ln 2 + \frac{L^3}{R^3} \Big).
  \label{coefP2}
\end{eqnarray}
Since the sign of Eq.~(\ref{coefP2}) changes at $L=(2\ln2/\pi)^{1/3}R$, if the phase transition is of second order the critical value of $L$ is given by this value. As in the text, this assumption is supproted numerically.

  Let us consider the fate of the second-order transition line that separates $\phi_x=0$ and $\phi_x\ne0$ (the red line in Fig.~\ref{fig:PhaseSU2}) in the limit $(L_\tau,L_x)\to(0,\infty)$. Assuming the second-order phase transition, this line is given by
  \begin{eqnarray}
    \frac{\partial^2 f}{\partial \psi_x^2}\Big|_{\psi_x=0}
    = \frac{\partial^2 f_{\rm pert}}{\partial \psi_x^2}\Big|_{\psi_x=0}
    + \frac{\partial^2 f_{\rm pot}}{\partial \psi_x^2}\Big|_{\psi_x=0} = 0 \ .
    \label{f(2)=0}
  \end{eqnarray}
  Since $P_\tau=1$ and thus $\psi_\tau=\pi/2$ is satisfied in this limit, the asymptotic form of $\partial^2 f_{\rm pert} / \partial \psi_x^2|_{\psi_x=0}$ is calculated to be
  \begin{eqnarray}
    &&\frac{\partial^2 f_{\rm pert}}{\partial \psi_x^2}\Big|_{\psi_x=0}
    = \frac{32}{\pi^2 L_\tau^4} \sum_{l_\tau,l_x=1}^\infty \frac{(-1)^{l_x} l_x^2}{ (l_\tau^2+l_x^2/r^2)^2}
    \nonumber \\
    &&= \frac8{L_\tau^4} \sum_{l_x=1}^\infty (-1)^{l_x}\Big( r^2 \sinh^{-2}\frac{l_x\pi}r + \frac{r^3}{\pi l_x} \coth \frac{l_x\pi}r -\frac2{\pi^2} \frac{r^4}{l_x^2} \Big)
    \nonumber \\
    && \xrightarrow[r\to0]{} \frac8\pi \frac1{L_\tau L_x^3} \sum_{l_x=1}^\infty \frac{(-1)^{l_x}}{l_x} = -\frac8\pi \ln(2) \frac1{L_\tau L_x^3} \ ,
  \end{eqnarray}
  with $r=L_\tau/L_x$.
  Since $\partial^2 f_{\rm pot} / \partial \psi_x^2|_{\psi_x=0} = 2/L_xR^3$,
  Eq.~(\ref{f(2)=0}) is satisfied at
  \begin{eqnarray}
    L_\tau = \frac{4\ln(2)}\pi \frac{R^3}{L_x^2} \simeq 0.8825 \frac{R^3}{L_x^2} \ .
    \label{crit}
 \end{eqnarray}

Next, we focus on the free energy for $N=3$ given by Eqs.~(\ref{FPertSU3}) and~(\ref{FHaarSU3}). Changing the variables as
\begin{eqnarray}
\psi_\tau \equiv \frac{2\pi}{3}-\phi_\tau\ , \ \ \psi_x \equiv \frac{2\pi}{3}-\phi_x
\end{eqnarray}
such that $\psi_c=0$ corresponds to $P_c=0$, the free energy reads
\begin{eqnarray}
f &=&\frac{8\pi^2}{405L_\tau^4}-\frac{2}{3L_\tau^4}\psi_\tau^2 - \frac{2}{3\pi L_\tau^4}\psi_\tau^3 + \frac{3}{2\pi^2 L_\tau^4}\psi_\tau^4 \nonumber\\
&+& \frac{8\pi^2}{405L_x^4} - \frac{2}{3L_x^4}\psi_x^2 - \frac{2}{3\pi L_x^4}\psi_x^3 + \frac{3}{2\pi^2 L_x^4}\psi_x^4  \nonumber\\
&+& \cdots \ .
\end{eqnarray}
That is, the free energy always includes cubic terms of $\psi_\tau$ and $\psi_x$with negative coefficients, generally resulting in a first-order phase transition. It is easily shown that the negative cubic terms also appear for $N\geq 3$. For this reason we infer that the first-order phase transitions and phase diagrams similar to Fig.~\ref{fig:PhaseSU3} are universally obtained for $N\geq 3$.

\bibliography{reference}

\end{document}